\documentclass[%
 aip,
 amsmath,amssymb,
 reprint,%
]{revtex4-1}

\usepackage{graphicx}% Include figure files
\usepackage{dcolumn}% Align table columns on decimal point
\usepackage{bm}% bold math
\usepackage{float} 
\makeatletter
\let\newfloat\newfloat@ltx
\makeatother

\usepackage{algorithm}
\usepackage{algpseudocode}
\usepackage[utf8]{inputenc}
\usepackage[T1]{fontenc}
\usepackage{mathptmx}
\usepackage{etoolbox}
\usepackage[english]{babel}

\usepackage[letterpaper,top=2cm,bottom=2cm,left=3cm,right=3cm,marginparwidth=1.75cm]{geometry}

\usepackage{amsmath}
\usepackage{graphicx}
\usepackage[colorlinks=true, allcolors=blue]{hyperref}
\usepackage{svg}

\usepackage{tikz}
\usepackage{braket}
\usepackage{amssymb}
\usepackage{subfigure}
\usepackage{comment}
\usepackage{hyperref}
\usepackage{listings}
\usepackage{mathtools,amssymb,bm}
\usepackage{booktabs}
\usepackage{psfrag}
\usepackage{amsthm}
\usepackage{epstopdf}
\usepackage{graphicx}
\usepackage{enumitem}
\usepackage{latexsym}
\usepackage{setspace}
\usepackage{fullpage}
\usepackage{tabularx}
\providecommand{\mathbold}[1]{\bm{#1}}
\newcommand{\mtx}[1]{\mathbold{#1}}

\usepackage[percent]{overpic}

\makeatletter
\def\@email#1#2{%
 \endgroup
 \patchcmd{\titleblock@produce}
  {\frontmatter@RRAPformat}
  {\frontmatter@RRAPformat{\produce@RRAP{*#1\href{mailto:#2}{#2}}}\frontmatter@RRAPformat}
  {}{}
}%
\makeatother
\begin{document}

\preprint{AIP/123-QED}

\title{Quantifying The Complex Spatiotemporal Chaos of Cardiac Fibrillation in  Ionic Models Across Parameter Regimes}
% Force line breaks with \\
\author{Xiaodong An}
 \affiliation{%
 School of Physics, Georgia Institute of Technology, Atlanta, Georgia 30332, USA}%%Lines break automatically or can be forced with \\
\author{Mikael Toye}%
% \email{Second.Author@institution.edu.}
\affiliation{%
 School of Physics, Georgia Institute of Technology, Atlanta, Georgia 30332, USA}%
\author{Abouzar Kaboudian}%
% \email{Second.Author@institution.edu.}
\affiliation{%
 School of Physics, Georgia Institute of Technology, Atlanta, Georgia 30332, USA}%

\author{Flavio H. Fenton}% 
\affiliation{%
 School of Physics, Georgia Institute of Technology, Atlanta, Georgia 30332, USA}%
 
\date{\today}% It is alsiways \today, today,

\begin{abstract}
Quantifying the complexity of cardiac systems is fundamental to understanding the onset of rhythm disorders, from mild arrhythmias to life-threatening fibrillation. In this work, for the first time, we develop a comprehensive pipeline to investigate how chaos appears and evolves in simplified cardiac models by calculating the largest Lyapunov exponent (LE) across different parameter sets. We show that both temporal and spatial LE ($\lambda^T$ and $\lambda^S$) estimators can be effectively used with action potential duration (APD) data, even without full access to state variables. Specifically, the spatial-temporal algorithm and Wolf's algorithm are used to quantify LEs as a measure of complexity using APD series drawn from various single-spiral patterns. We also identify the minimum data length and spatial density necessary to achieve robust and accurate LE estimation. Our results suggest that these APD-based methods can be applied not only to simulation data but also to clinical or experimental data, particularly when observations are limited, such as when only APD data are available for analysis.
\end{abstract}

\maketitle

\begin{quotation}
Irregular cardiac activities, such as arrhythmia and fibrillation, are often linked to chaotic electrical activity in the heart. Measuring this chaos typically requires full access to state variables that are hard to obtain in real-world settings. In this study, we show that a simple and commonly measured signal, APD, can be used to reveal key features of cardiac instability, offering a practical tool for both experimental and clinical applications.
\end{quotation}

\section{Introduction}
In recent years, there has been growing interest in understanding and characterizing spatiotemporal chaos in cardiac tissue~\cite{mulimani2022dmd,molavi2022spatio,loppini2018spatiotemporal}. Spatiotemporal chaos can emerge in cardiac tissue due to its inherent nonlinear dynamics, causing dangerous cardiac rhythm disorders such as fibrillation~\cite{1,2,garfinkel1997,gray1998}. Being able to quantify the chaos could deepen our understanding of cardiac instability and guide early diagnosis and treatment strategies~\cite{weiss2006,jalife2000}. While the largest Lyapunov exponent (LE) is widely used as an indicator for chaos, most studies focus on temporal LE ($\lambda^T$) derived from full voltage data~\cite{rosenstein1993,5}. In practical scenarios, however, observed quantities may not be state variables but rather may be derived values, such as action potential durations (APDs) from optical mapping or electrocardiograms (ECGs), and may be limited to short recordings. Also, spatial aspects of dynamics are often overlooked in conventional approaches~\cite{hagerman1996,casaleggio1997,valenza2012,pavlov2000}. As a result, there is a need for resource-efficient methods that can take spatial information into account.

This study aims to quantify spatiotemporal chaos from simulations of ionic cardiac cell models and explore whether APD sequences in time and space can be used to quantify the underlying dynamics of the system, and to provide a measurement for chaos as parameters of the models are varied.  We apply both temporal LE (Wolf’s Algorithm~\cite{5}) and spatial-temporal LE~\cite{7} (SLE) methods to data obtained from ionic models, including the Fenton-Karma (FK) model~\cite{1} and Ten Tusscher-Noble-Noble-Panfilov (TNNP) human ventricular model~\cite{ten2006alternans}, for various parameter sets that produce different dynamics. Through a systematic evaluation of several single-spiral wave patterns, we find that the two methods respond differently depending on whether the chaos originates from spiral tip drifting or from shape deformation. By comparing the LE results with parameter transitions, we can establish a link between chaotic transitions and model gating parameters.

Moreover, we conduct a series of experiments to identify how temporal length, spatial sampling resolution, APD measurement resolution and noise level affect the stability and accuracy of chaos quantification. We find that the temporal LEs calculated from APD signals ($\lambda^T(\text{APD}))$ require longer sequences but offer higher precision, while the spatial LEs calculated from APD signals ($\lambda^S(\text{APD}))$  converge faster with fewer data points. Furthermore, when limited spatial points are available, utilizing random sampling allows $\lambda^T(\text{APD})$ to achieve a smaller margin of error. Regarding APD measurement resolution, both methods yield similar outcomes and require a minimum measuring step of $dt = 20$. We also investigate the effect from Gaussian noise, and show that both methods remain robust at Gaussian noise levels of up to 10\%. These findings are useful for real-world applications, especially when dealing with noisy or short-duration experimental recordings. Our goal is to bridge theoretical modeling with practical application by showing that LE-based chaos estimation could be extended to experimental and machine learning pipelines using APD sequences alone.

\section{Methods}

\begin{figure*}
\centering
\includegraphics[width=1.0\linewidth]{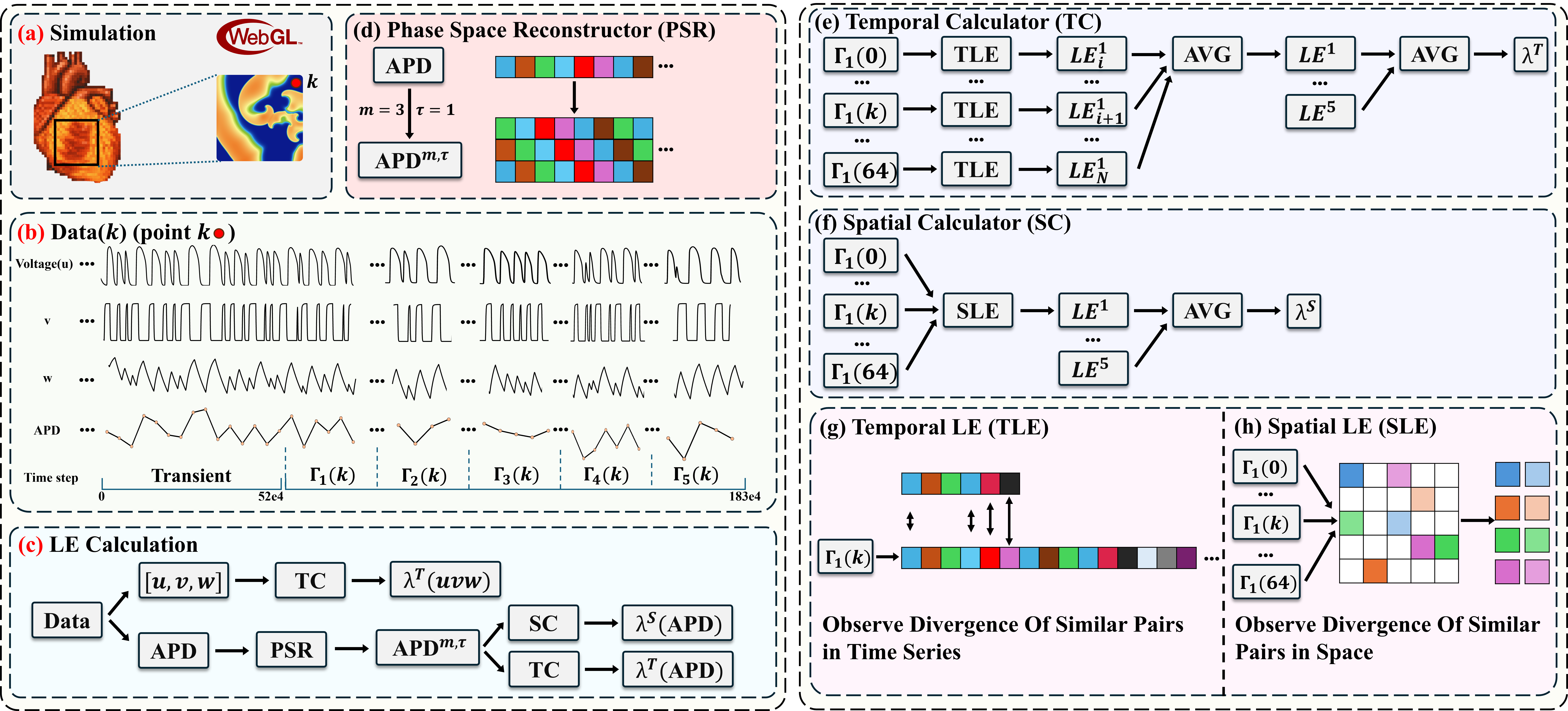}
\caption{Schematic overview of the methodology for estimating LEs in cardiac systems, exemplified using the FK model with the multiple spiral wave pattern. Panels (a)–(c) (highlighted in red) illustrate the primary workflow, while (d)–(h) provide detailed descriptions of the underlying computational components and algorithmic implementations. (a) Simulation: The FK model is applied to simulate cardiac electrical activity using WebGL acceleration. A specific measuring point i is selected for later demonstration. (b) Data$_i$:  Time series are recorded for the three state variables and the calculated APD. APD is calculated for later steps. An initial transient period of 520~s is skipped, and the remaining measurement time is divided into 5 zones ($\Gamma_1(\mtx k)$ to $\Gamma_5(\mtx k)$). (c) LE Calculation: Data is used in two separate ways: as a full set of state variables and as a partial observation where only APD is known. (d) Phase Space Reconstructor (PSR): With proper embedded dimension and lagging, APD could reconstruct the original phase space by lag-embedding, shown as $\text{APD}_{m,\tau}$. The APD sequence is chromatically encoded as a schematic abstraction of its temporal trajectory. (e) Temporal Calculator (TC): For each region, Temporal LE (TLE) is calculated at different points and then averaged to obtain the final $\lambda^T$. (f) Spatial Calculator (SC): For each region, all measuring points are used inside the algorithm to compute one final spatial LE (SLE). (g) TLE: Based on Wolf's algorithm, TLE is calculated by observing the divergence of similar pairs within the time series (h) SLE: Using the spatial-temporal algorithm, all measuring points in the tissue are utilized to calculate the SLE by observing divergence of similar pairs in space.}
\label{fig:flowchart}
\end{figure*}

LE is a quantitative measure of the divergence rate for nearby trajectories, implying the stability across both linear and nonlinear systems in spatial or temporal domains. Typically, following that basic definition, in a system with single state variable $x$, the LE can be naively calculated as \cite{5}:

\begin{equation}
    \lambda(x_0) = \lim_{t\rightarrow\infty}\lim_{\delta x_0 \rightarrow 0} \frac{1}{t}\ln{\frac{\delta x_t}{\delta x_0}}.
\end{equation}

where $\delta x_t = x_t - x'_t$ is the difference at time $t$ between a reference system and a second system starting from the perturbed initial condition $x'_0 = x_0 + \delta x_0$.

Now, heading to the system with $n$ state variables, it can be computed from linearization \cite{11}:
\begin{equation}
    \lambda(\mtx X_0) = \lim_{{t \to \infty}} \frac{1}{t} \ln \frac{\| J^t(\mtx X_t) \delta \mtx X_0 \|}{\| \delta \mtx X_0 \|},
\label{Eq LE}
\end{equation}

where:

$\bullet$ $\mtx{X}_0 = (x^1_0, x^2_0, \dots, x^n_0)$ represents the state vector at $t = 0$,

$\bullet$ $J^t(\mtx X_t) = \Pi_{t' = 1}^{t' = t} J(\mtx X_t')$,

$\bullet$  $J(\mtx{X}_{t}) = \begin{bmatrix}
\frac{\partial \mtx{X}_{t}}{\partial x^1_{t}}  \cdots \frac{\partial \mtx{X}_{t}}{\partial x^n_{t}}
\end{bmatrix}$ is the Jacobian matrix.

However, in practical cases where the Jacobian matrix is impossible to obtain (too many state variables or indeterminate original partial differential equations (PDEs) of the system), several data-driven algorithms to estimate LEs, including $\lambda^T$ and $\lambda^S$, have been developed. Here we utilized Wolf's algorithm\cite{5} for $\lambda^T$ estimation and the spatio-temporal algorithm by Sol\'e and Bascompte \cite{7} for $\lambda^S$ estimation, as described below.The overall pipeline of the methodology in this paper, including simulation, data acquisition, phase space reconstruction, and LE calculation, can be seen in Fig.~\ref{fig:flowchart}

\subsection{Calculating Temporal LE, $\lambda^T$}

Wolf's algorithm\cite{5} for calculating LE is widely utilized to quantify the rate at which nearby trajectories in a dynamical system diverge. Instead of relying solely on one trajectory with initial point $\mtx X_0$, as shown in Eq.~\ref{Eq LE}, the algorithm selects multiple trajectories in a temporal data set. This approach mitigates the risk of selecting a non-representative trajectory, such as one located on a periodic orbit or within atypical attractors, thereby ensuring a more robust and accurate estimation of the LE. However, the method does not incorporate spatial information if the points are correlated, such as by diffusive coupling in the case of cardiac tissue. 

To investigate the performance of different LE estimation methods, we applied the algorithm both to the full set of state variables $u$,$v$, and $w$ and to a more limited set of observations based on APD. For clarity, we refer to these as the $uvw$ (temporal) method and the APD (temporal) method, respectively. Specifically, the data input for these methods corresponds to 
$\{\mtx{X}_i = (u_i, v_i, w_i)\}, \; i = 1, \ldots, N$, with $N$ denoting the total number of time steps at a given point,
and
$\{\mtx{X}_i = \mathrm{APD}_i\}, \; i = 1, \ldots, N_{\mathrm{APD}}$, 
where $N_{\mathrm{APD}}$ denotes the number of APD values available at a given point, for the $uvw$ (temporal) and APD (temporal methods, respectively.

Also, for partial observation, which projects the full state of a dynamical system onto a few observable coordinates or a specific function of state variables, such as APD, it is known that a proper phase space reconstruction can recover the same LE as those computed from full state space observation \cite{8}. In the temporal context, at some point $k$, this equivalence in terms of LE between the original and reconstructed system can be expressed as
\begin{equation}
\begin{aligned}
    \mtx{\Gamma}^{m,\tau}(\mtx k) &\equiv \big\{\mtx {X}_i = (\mathrm{APD}_i)^{m,\tau}\big\}, \\
    \mtx{\Gamma}(\mtx k)          &\equiv \big\{\mtx {X}_i = (u_i, v_i, w_i)\big\}, \\
    \mtx{\Gamma}^{m,\tau}(\mtx k) &\cong \mtx{\Gamma}.
\end{aligned}
\label{Eq ps}
\end{equation}
where $\Gamma$ represents a temporal sequence and $(\mathrm{APD}_i)^{m,\tau}$ denotes the APD sequence properly embedded with dimension $m$ and time lag $\tau$.

In either case ($uvw$ or APD observation), the algorithm then proceeds by locating a neighbor for each point in the temporal trajectory that is close in state (both their magnitude and direction) but not in time to avoid correlations between temporally adjacent trajectories.

Specifically, a point $\mtx{X}_i = (u_i, v_i, w_i)$ (or $\mtx{X}_i = \mathrm{APD}_i$) where $i \approx N_{total}/2$ ($N_{total} = N$ if using uvw sequence or $N_{total} = N_{APD}$ if using APD sequence.) is chosen. We find its nearest neighbor $\mtx{X}_{j}$ as the point in the series that minimizes the Euclidean distance:
\begin{equation}
\begin{split}
    L_{ij} &= \lVert \mtx{X}_{i} - \mtx{X}_{j} \rVert
    \\
     \mathop{\arg \min}\limits_{j}L_{ij} &< \epsilon,
\label{Eq pair}
\end{split}
\end{equation}
where $j \neq i$, $\frac{ \mtx{X}_{i} \mtx{X}_{j} }{ |\mtx{X}_{i}|| \mtx{X}_{j}|}< \theta$, $\theta = \frac{\pi}{9}$ is the maximum initial angular distance, as suggested by Wolf et al.~\cite{5} and $\epsilon$ is the maximum initial separation.

Then  $L_{ij}$ is evolved by one time step each until
\begin{equation}
    L_{ij} \geq \epsilon \text{ or } i \geq N \text{ or } j\geq N.
\label{Eq evolve}
\end{equation}
Currently, there is no universal standard for selecting the optimal maximum initial separation $\epsilon$, also referred to as the neighborhood search radius. Previous studies\cite{18,sano1985measurement,eckmann1986liapunov} suggest that $\epsilon$ must be large enough to exceed the data’s noise floor and ensure a sufficient density of neighbors for a statistically valid calculation. The $\epsilon$ must also remain small enough to stay within the attractor's local regime. An excessively large $\epsilon$ may lead to a saturation of neighbors, significantly increasing computational cost and obscuring the underlying chaotic dynamics.

Following these principles and the numerical test of our algorithm, we selected $\epsilon = 0.8$ for the $uvw$ sequence. This value provides a robust balance between neighbor count, computational efficiency, and a LE within a reasonable $[0, 1]$ range. For the APD sequence, we used grid search to set $\epsilon \approx \overline{\text{APD}_{\text{all}}}$, where $\overline{\text{APD}_{\text{all}}}$ denotes the average value calculated across all data points measured in the FK model, ensuring that its LE remains comparable to that of the $uvw$ sequence for consistent analysis. This same criteria is applied to the spatial LE algorithm discussed in later sections.

Then record $L_{ij}$ and repeat the evolution in Eq.~\ref{Eq evolve} until $i\geq N$

By recording the growth ratios $L_{i'}/L_i$ across all evolutions, the temporal LE is calculated as \cite{5}
\begin{equation}
    \lambda^T (\mtx \Gamma(\mtx k)) \equiv \frac{1}{i_f - i_0}\sum_{\text{All Loops}}\log_2\frac{L_{i'}}{L_i}.
\label{Eq wolf LE}
\end{equation}

More details could be seen in the pseudocode in Algorithm~\ref{TLE codes}.

In the single-spiral wave cases discussed later, a time series is recorded for every pixel in the tissue, each having a corresponding $\lambda^T(\mtx \Gamma(\mtx k))$. To obtain a representative measure for the system, these values are averaged across the domain as follows:
\begin{equation}
    \lambda^T_{\text{single spiral}} = \frac{1}{64}\sum_{\mtx k = \text{point}1}^{\text{point}64}\lambda^T(\mtx \Gamma(\mtx k)).
\label{Eq TLE_single_spiral}
\end{equation}

While for the multiple spiral wave cases, it exhibits higher complexity and requires a longer observation to ensure robust estimation. we used a longer measurement. To maintain algorithmic efficiency and accuracy, the total time series $\mtx \Gamma$ is partitioned into five distinct zones, $\mtx \Gamma_1$ through $\mtx \Gamma_5$, as shown in Fig.~\ref{fig:flowchart}(b). The resulting temporal LE is calculated by averaging across both these temporal zones and the spatial measurement points:
\begin{equation}
    \lambda^T_{\text{multi spiral}} = \frac{1}{5}\frac{1}{64}\sum_{z=1}^{5}\sum_{\mtx k = \text{point}1}^{\text{point}64}\lambda^T(\mtx \Gamma_z(\mtx k)).
\label{Eq TLE_multi_spiral}
\end{equation}

\begin{algorithm}
\caption{Temporal LE ($\lambda^T$)}
\begin{algorithmic}[1]
\State \textbf{Input:} Time series $\mtx{\Gamma} = \{\mtx{X}_1, \mtx{X}_2, \dots, \mtx{X}_{N_{total}}\}$
\State \textbf{Data Configuration:} 
\State \quad \textbf{if} using full state variables: 
\State \qquad $\mtx{X}_i = (u_i, v_i, w_i)$, $N_{total} = N$
\State \quad \textbf{else if} using APD sequences:
\State \qquad $\mtx{X}_i = (APD_i, \dots, APD_{i+(m-1)\tau})$, $N_{total} = N_{APD}$
\State \textbf{Parameters:} Max initial angular distance $\theta = \pi/9$, Max initial separation $\epsilon$
\State \textbf{Initialize:} $i \gets \lfloor N_{total}/2 \rfloor$, $i_0 \gets i$, $SumLog \gets 0$

\While{$i < N_{total}$}
    \State \textbf{1. Nearest Neighbor Search:} 
    \State Find $j$ minimizing $L_{ij} = \lVert \mtx{X}_i - \mtx{X}_j \rVert$ such that $j \neq i$, $L_{ij} < \epsilon$, and $\frac{\mtx{X}_i \cdot \mtx{X}_j}{\lVert\mtx{X}_i\rVert\lVert\mtx{X}_j\rVert} < \theta$   
    \State \textbf{2. Trajectory Evolution:}
    \State $i' \gets i, j' \gets j$
    \State $L_{i'j'} \gets \lVert \mtx{X}_{i'} - \mtx{X}_{j'} \rVert$
    \While{$L_{i'j'} < \epsilon$ \textbf{and} $i' < N_{total}$ \textbf{and} $j' < N_{total}$}
        \State $i' \gets i' + 1, j' \gets j' + 1$
        \State $L_{i'j'} \gets \lVert \mtx{X}_{i'} - \mtx{X}_{j'} \rVert$
    \EndWhile
    
    \State \textbf{3. Accumulate and Iterate:}
    \State $SumLog \gets SumLog + \log_2 \frac{L_{i'j'}}{L_ij}$ 
    \State $i \gets i'$ 
\EndWhile
\State $i_f \gets i$
\State \textbf{Output:} 
$\lambda^T = \frac{1}{i_f - i_0} SumLog$
\end{algorithmic}
\label{TLE codes}
\end{algorithm}

\subsection{Calculating Spatial LE, $\lambda^S$}

Classical LE algorithms like Wolf's Algorithm mainly focus on temporal chaos. In contrast, SLE can quantify how local perturbations evolve and spread through space over time, which provides a more comprehensive understanding of the global dynamics of the system by capturing both temporal and spatial variations and instabilities. In cardiac systems, the state variables typically evolve over a cardiac cycle and therefore do not diverge significantly within a single step. However, as will be shown later, one-step divergence forms the core idea of the method, so here we use APD alone, with $\mtx{X}$ in the general exposition of the method below later implemented as APD.

For the SLE method, a 2D map $\mtx \Lambda^2(L)$ is defined as
\begin{equation}
    \mtx \Lambda ^2(L) = \{\mtx{k} = (\alpha, \beta) \mid 1 \leq \alpha,\beta \leq L\},
\end{equation}
where $L$ is the length of the map,
$\mtx k$ is a pixel in the map, and
$\alpha,\beta$ are the $x,y$ coordinates of the pixel $\mtx k$.

It can be seen that
\begin{equation}
    \mtx{\Gamma}^{m,\tau}(\mtx \Lambda^2) \cong \mtx{\Gamma}(\mtx \Lambda^2) = \bigcup_{\mtx k \in \mtx \Lambda} \mtx{\Gamma}(\mtx{k}).
\end{equation}\\

For each $\mtx{X}_i(\mtx{k}) \in \mtx{\Gamma}(\mtx{k}), \forall \mtx k \in \mtx \Lambda^2$, we search its neighbors $\mtx h \in \mtx \Lambda^2, \mtx h \neq \mtx k$ to identify pairs where the following condition holds:
\begin{equation}
    L_{i}(\mtx{k},\mtx{h}) = \lVert \mtx{X}_{i}(\mtx{k}) - \mtx{X}_{i}(\mtx{h})  \rVert < \epsilon,
\label{eq 9}
\end{equation}
where $\epsilon$ is the maximum initial separation. Here, the maximum initial separation is established as $\epsilon = \overline{\text{APD}}/2$.
Then, each pair identified as being sufficiently close at iteration $i$ is evolved one more step by calculating
\begin{equation}
    L_{i+1}(\mtx{k},\mtx{h}) = \lVert \mtx{X}_{i + 1}(\mtx{k}) - \mtx{X}_{i + 1}(\mtx{h})  \rVert.
\label{Eq 10}
\end{equation}\\
Finally, the SLE is defined as
\begin{equation}
\begin{split}
    \lambda^S(\mtx{\Gamma}^{m,\tau}(\mtx \Lambda^2)) = &\frac{1}{\text{\# of Pair}}\sum_{i = 1}^{N_{total}-1 }\sum_{\braket{\mtx k,\mtx h}} 
    \\
    &\log_2(L_{i+1}(\mtx{k},\mtx{h})/L_{i}(\mtx{k},\mtx{h})),
\end{split}
\label{eq 11}
\end{equation}

where the second summation is taken over all pairs that satisfy Eq.~\ref{eq 9}. Detailed pseudocode is available in Algorithm~\ref{SLE codes}.

A key distinction of the spatial LE method is that it inherently incorporates all measurement points and thus no additional spatial averaging is required. For the single and multiple spiral wave cases analyzed in this study, we have:
\begin{equation}
\lambda^S_{\text{single spiral}} = \lambda^S(\mtx{\Gamma}^{m,\tau}(\mtx \Lambda^2)).
\label{Eq SLE_single_spiral}
\end{equation}

\begin{equation}
\lambda^S_{\text{multi spiral}} = \frac{1}{5}\sum_{z=1}^{z=5}\lambda^S(\mtx{\Gamma}_z^{m,\tau}(\mtx \Lambda^2)).
\label{Eq SLE_multi_spiral}
\end{equation}

For simplicity of expression in the following sections, we will omit the "single spiral" and "multi spiral" subscripts. Instead, $\lambda^T$ and $\lambda^S$ will be used to represent these respective temporal and spatial LEs throughout the remainder of the analysis.

\begin{algorithm}
\caption{Spatial LE ($\lambda^S$)}
\begin{algorithmic}[1]
\State \textbf{Input:} Spatio-temporal field $\mtx{\Gamma}(\mtx \Lambda^2) = \{\mtx{X}_i(\mtx{k}) \mid 1 \le i \le N_{total}, \mtx{k} \in \mtx \Lambda^2\}$
\State \textbf{Data Configuration:} 
\State \quad $\mtx{\Lambda}^2(L) = \{\mtx{k} = (\alpha, \beta) \mid 1 \le \alpha, \beta \le L\}$
\State \quad $\mtx{X}_i(\mtx{k}) = (APD_i(\mtx{k}), \dots, APD_{i+(m-1)\tau}(\mtx{k}))$
\State \textbf{Parameters:} Max initial separation $\epsilon = \overline{\text{APD}}/2$, $L = 8$, $N_{total} = N_{APD}$
\State \textbf{Initialize:} $SumLog \gets 0$, $N_{pair} \gets 0$

\State \textbf{1. Spatial Neighbor Search:} 
\For{each pixel $\mtx{k} \in \mtx \Lambda^2$}
    \For{each neighbor $\mtx{h} \in \mtx \Lambda^2, \mtx{h} \neq \mtx{k}$}
    
        $N_{total} \gets min(N_{APD(\mtx{k})},N_{APD(\mtx{h})})$
        \For{$i = 1$ \textbf{to} $N_{total}-1$}
            \State $L_{kh, i} \gets \lVert \mtx{X}_{i}(\mtx{k}) - \mtx{X}_{i}(\mtx{h}) \rVert$
            
            \If{$L_{kh, i} < \epsilon$}
                \State \textbf{2. One-Step Evolution:}
                \State $L_{kh, i+1} \gets \lVert \mtx{X}_{i+1}(\mtx{k}) - \mtx{X}_{i+1}(\mtx{h}) \rVert$
                
                \State \textbf{3. Accumulate:}
                \State $SumLog \gets SumLog + \log_2 \left( \frac{L_{kh, i+1}}{L_{kh, i}} \right)$
                \State $N_{pair} \gets N_{pair} + 1$
            \EndIf
        \EndFor
    \EndFor
\EndFor

\State \textbf{Output:} 
$\lambda^S = \frac{1}{N_{pair}} SumLog$
\end{algorithmic}
\label{SLE codes}
\end{algorithm}

\subsection{Phase Space Reconstruction}

Rather than solely quantifying the original phase space, we also investigated the scenario involving a restricted observation of state variables, where, for example, APD is used as the partial observable. We applied phase space reconstruction on APD, which is based on Takens' Embedding Theorem\cite{8}. For Takens' embedding, two variables need to be determined: embedding dimension $m$ and lag $\tau$. The lag $\tau$ is typically selected first, followed by the embedding dimension $m$.

Embeddings with the same $m$ but different $\tau$ are equivalent in the mathematical sense for noise-free data\cite{18}. Therefore, for the purposes of this research, where simulations are conducted without the influence of noise, a simple choice of $\tau = 1$ is sufficient.

The embedding dimension refers to the number of delayed elements for the time series $\gamma = \{x_i\}$ used to construct the state space $\mtx\Gamma = \{\mtx X_i\}$. To reconstruct phase space successfully, the sufficient condition would be\cite{8},\cite{22}
\begin{equation}
    m > 2d, m > 2d_c,
\end{equation}
where $d$ is the phase space dimension and $d_c$ is the box-counting dimension of strange attractors.
However, Sugihara et al.~\cite{23} state that the embedding dimension can be
   $ m \approx d$,
and they successfully reconstructed the Lorenz system ($d = 3$) with $m = 3$ by plotting the reconstructed phase space and original phase space and showing that they are geometrically the same.
Therefore, we may conclude that $m$ in $[d,2d]$ or $[d,2d_c]$ could be appropriate, with its exact minimum value depending on the nature of the dataset being analyzed. Several methods exist for determining an optimal embedding dimension, such as the False Nearest Neighbor (FNN) \cite{24}, the Characteristic Decay Rate \cite{7}, the Fuzzy Clustering \cite{25}, the Fill-Factor method \cite{26}, and the Average Integral Local Deformation \cite{26} methods. A comprehensive summary of these algorithms can be found in Jiang and Adeli \cite{25}. Most of the approaches above rely on the fact that, once $m$ exceeds the minimum embedding dimension, the reconstructed phase space would be geometrically the same as the true phase space. Furthermore, a reasonable data length required to determine the embedded dimension is at least $10^{m}$ \cite{27}. Accordingly, the success of those algorithms requires the dynamical system to behave in a low dimension with sufficiently long data as input.

In this paper, we apply the FNN method with $\tau = 1$. We first initialize different embedded dimensions $\{m\}$.
For every embedded vector with a certain $m$: 
\begin{equation}
\begin{split}
   \mtx{X}_i^{m,\tau = 1} = \{x_i, x_{i+\tau}, x_{i+2 \tau}, \ldots, x_{i+(m-1)\tau}\}, \\i = 1, 2, \ldots, N -m\tau + \tau,
\end{split}
\end{equation}
we find its nearest neighbor $\mtx{X}_j^{m}$ such that $j = \mathop{\arg \min}\limits_{j}\lVert\mtx{X}_i^{m}  - \mtx{X}_j^{m} \rVert = \mathop{\arg \min}\limits_{j}{R}_i^{m}$. 

We obtain ${R}_i^{m+1} = \lVert{\mtx X}_i^{m+1}  - {\mtx X}_j^{m+1} \rVert$. Te pair $i,j$ will be identified as false neighbors if $| {R}_i^{m} - {R}_i^{m+1} | \gg 0 $.
Specifically, if the ratio
%\begin{equation}
 $    R(i,m) = \sqrt{1-({R}_i^{m+1})^2 /({R}_i^{m})^2}$
%\end{equation}
is larger than a pre-defined value $R_0 = 10$, we call $\{i,j\}$ a pair of false neighbors.
%or point $i$ has a false neighbor $j$.

Consequently, if at a certain embedding dimension $m_0$ the percentage of nearest neighbors that are false 
%false nearest neighbors $\frac{\text{false pairs of neighbors}}{\text{total pairs of neighbors}}$ 
stabilizes and does not decrease with a further increase in the embedding dimension, we can identify $m_0$ as the appropriate dimension.

\subsection{FK Model}

%The FK model was developed in 1990s and can quantitatively reproduce the APD of a cell as a function of its diastolic interval (DI), APD vs DI curves (restitution curves) are used to quantiatively determine the dynamics of cardiac cells as their period of excitation changes\cite{1}. This model's aims are to provide a simplified solution to simulate the electrical activity in cardiac tissue. Moreover, it balances the computational cost and physiological accuracy, making it quite useful in large-scale simulations with studies of arrhythmic behaviors in the heart.

The FK model describes the rate of change of the membrane potential of a cardiac cell using three transmembrane currents:  fast inward ($I_{\text{fi}}$), slow inward ($I_{\text{si}}$), and slow outward ($I_{\text{so}}$) currents. The model's three state variables are defined as the membrane voltage ($u$) along with two gating variables ($v,w$) that control the inactivation and reactivation of currents.
The FK model equations are given as \cite{2}
\begin{equation}
\begin{split}
    \frac{\partial u(\mtx{x},t)}{\partial t} &= D\nabla^2 u - \frac{I_{\text{fi}}(u,v) + I_{\text{so}}(u) + I_{\text{si}}(V,w))}{C_{m}}
\\
    \frac{\partial v(\mtx{x},t)}{\partial t} &= \frac{(1 - \mathcal{H}(u-u_c))(1 - v)} {\tau_v^-(u)} - \frac{\mathcal{H}(u-u_c)v}{\tau_v^+}
\\
    \frac{\partial w(\mtx{x},t)}{\partial t} &= \frac{(1 - \mathcal{H}(u-u_c))(1 - w)}{\tau_w^-} - \frac{\mathcal{H}(u-u_c)w}{\tau_w^+},
\label{FK Voltages}
\end{split}
\end{equation}
where the three currents are specified as
\begin{equation}
\begin{split}
    I_{\text{fi}}(u,v) &= -v\mathcal{H}(u-u_c)(u - u_c)(1 - u) / \tau_d
\\
    I_{\text{so}}(u) &= u(1 - \mathcal{H}(u-u_c)) / \tau_0 + \mathcal{H}(u-u_c) / \tau_r
\\
    I_{\text{si}}(u,w) &= -w(1 + \textrm{tanh}(k(u - u_c^{\text{si}}))) / (2\tau_\text{si})
\label{FK Current}
\end{split}
\end{equation}
with $ \tau_v^-(u) = (1-\mathcal{H}(u-u_v))\tau_{v1}^- - \mathcal{H}(u-u_v)\tau_{v2}^-$
and $\mathcal{H}(x)$ the Heaviside function.

\subsection{TNNP Human Model}

To evaluate the robustness of our LE estimation method beyond the simplified ionic model, we also use the TNNP human ventricular model, which exhibits higher physiological complexity, with 19 state variables per cell and 12 ionic currents governing the action potential dynamics~\cite{ten2006alternans}. This model is based on experimental measurements of human APD restitution and includes a more extensive description of intracellular calcium dynamics.

The TNNP model is described by the following equation:
\begin{equation}
\begin{split}
\frac{\partial u(\mtx{x},t)}{\partial t} = D\nabla^2 u - \frac{1}{C_{m}}(I_\text{Na} + I_\text{bNa} + I_\text{CaL} + I_\text{bCa} + I_\text{pCa} \\+ I_\text{Kr} + I_\text{Ks} + I_\text{K1} + I_\text{to} + I_\text{pK} + I_\text{NaCa} + I_\text{NaK}).
\\
\label{TNNP Equ}
\end{split}
\end{equation}

While the governing equation for voltage is structurally analogous to the FK model, the multiple currents and state variables result in more intricate dynamics. This elevated level of complexity makes the TNNP model a good testing case for the robustness of our LE estimation method. 
%By challenging the algorithm with a more complex and realistic physiological model, we could better assess its utility in clinical or experimental settings.

\section{Data Acquisition}

\begin{table}[htbp]
\centering
\caption{Simulation configuration details, including temporal and spatial discretizations, texture size, number of measurement pixels, and time intervals for transient removal and data collection.}
\vspace{0.5em}
\begin{tabular}{lcc}
\hline\hline
\textbf{Parameter} & \textbf{Symbol} & \textbf{Value} \\
\hline
Simulation time step (FK)     & $d\tau$     & 0.1 ms\\
Simulation time step (TNNP)     & --    & 0.02 ms\\
Measurement time step     & $dt$       & 4 ms\\
Space step (FK)              & $dx$, $dy$ & 0.035 cm\\
Space step (TNNP)              & -- & 0.025 cm\\
Texture size (FK)             & --         & $512 \times 512$ \\
Texture size (TNNP)             & --         & $1000 \times 1000$ \\
Domain size (FK)             & --         & 18 cm \\
Domain size (TNNP)             & --         & 25 cm \\
Measurement pixels        & $\{\mtx k\}$ & $8 \times 8$ \\
\hline\hline
\end{tabular}
\begin{flushleft}
%\small $^*$ Units in cm
\end{flushleft}
\label{tab:simulation_parameters}
\end{table}

\begin{figure*}
    \centering
    \subfigure[]{\includegraphics[width=0.2\textwidth]{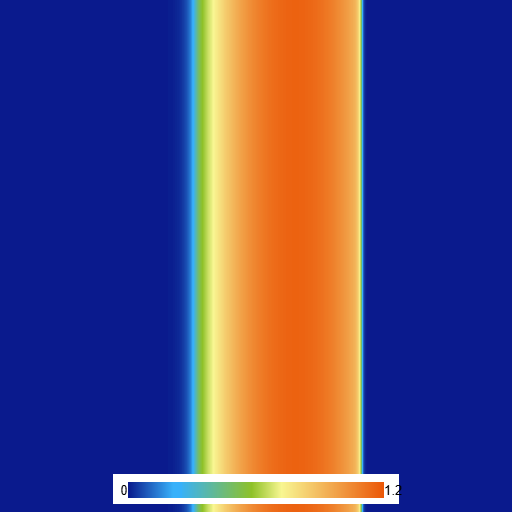}} 
    \subfigure[]{\includegraphics[width=0.2\textwidth]{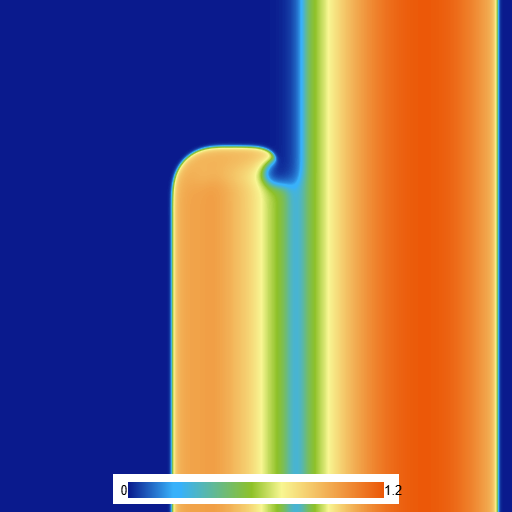}} 
    \subfigure[]{\includegraphics[width=0.2\textwidth]{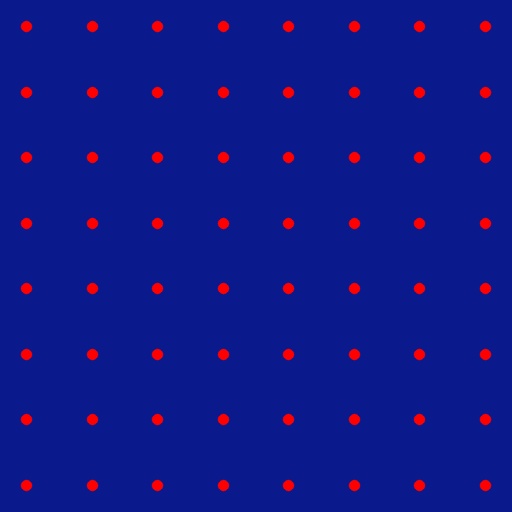}}\\
    \caption{Spiral wave initiation and spatial sampling points for LE analysis. 
    %Spiral waves and possible fibrillation are initiated for all parameter sets studied.  
    (A) Propagating wave originated by a stimulus (S1) applied along the  left edge of the tissue.  (B) Spiral wave initiated from a second stimulus (S2) applied behind the first wave in the lower half of the tissue. (C) Example of spatial sampling points used for LE calculation, indicated by the red dots ($8 \times 8$ grid).}
\label{fig:initial condition}
\end{figure*}

Voltage and APD data from the FK and TNNP human ventricular models were simulated in two-dimensional (2D) tissue using browser-based simulations. APD was calculated using a threshold of $u = 0.3$, where linear interpolation was applied between time steps to precisely determine the crossing points when the measurement time step did not land exactly on the threshold. The Diastolic Interval (DI) was calculated as the time elapsed between the end of one APD and the onset of the subsequent APD. These simulations were implemented with WebGL via the Abubu.js library, which provides a simplified interface for GPU acceleration compared to traditional WebGL. Abubu.js uses textures as the primary data structures, where each texture contains 512*512 pixels (1000*1000 pixels for TNNP model), as shown in Table~\ref{tab:simulation_parameters}, and every pixel has four channels: red ($r$), green ($g$), blue ($b$), and alpha ($\alpha$, transparency). In the FK model, only three channels ($r,g,b$) are used to store state variables ($u,v,w$) and $\alpha$ is set to be $1.0$. In the TNNP model, five separate textures are used, each storing at most four state variables (19 state variables in total per cell).

The PDEs were solved numerically using the forward Euler method for time integration and the second-order central difference scheme for the Laplacian. The time and space discretizations are shown in Table~\ref{tab:simulation_parameters}. No-flux boundary conditions were enforced.

We conducted simulations using multiple parameter sets to quantify the complexity of their dynamics, especially for spiral waves undergoing breakup. The same S1--S2 cross-field protocol~\cite{narayan2008repolarization} was used consistently across all configurations to initiate spiral activity. This protocol applies an initial planar stimulus (S1), followed by a carefully timed second stimulus (S2) perpendicular to S1 during the refractory tail of the wavefront, facilitating the formation of spiral wave singularities, as shown in Fig.~\ref{fig:initial condition}(A,B), allowing subsequent nonlinearity calculation.

For the multiple spiral wave pattern in the FK model, we discarded an initial transient $T_t = 520$~s (approximately 1300 APDs) to ensure that the phase space trajectory settles onto one stable attractor or a group of attractors whose chaoticity we aim to quantify. This practice is consistent with common convention in cardiac simulations, where dynamics in the early stage are typically excluded to 
%eliminate transient behavior resulting from initialization. Discarding the transient 
ensure that the measured chaos reflects the long-term behavior of the attractor, not the initial evolution from stimulus onset. Following the transient, we collected data over a simulation window $\Delta T = 1,310$~s, which is long enough to capture the full geometry of the attractor and any potential transitions between distinct attractors. As shown in Fig.~\ref{fig:initial condition}(C), APD signals were recorded every 4~ms for both models
%40 $d\tau$ (200 $d\tau$ for TNNP Human model) 
on an $8\times 8$ uniform grid 
%at \( 8^2 \) evenly spaced spatial locations 
across the tissue.

To assess temporal variability in LE estimates, each post-transient simulation was divided into five equal-duration zones, and the LE was computed independently for each, as shown in Eq.~\ref{Eq TLE_multi_spiral} and Eq.~\ref{Eq SLE_multi_spiral}. The standard deviation of these values is shown as the envelope width of the shaded region.

The configuration for the TNNP model was adjusted due to its greater physiological complexity. A shorter transient of 10~s was utilized, as spiral waves in this model often cannot be sustained for a long duration. Consequently, the measurement region for the TNNP model was defined between 10 and 260~s.

Experiments assessing data quality were conducted using specific data configurations. To ensure a representative range of spatiotemporal chaos, these experiments employed FK model datasets generated from all four time constants ($\tau_d, \tau_0, \tau_r, \text{ and } \tau_{si}$). To evaluate the effect of the number of spatial points used,
%(as shown in Fig.~\ref{fig:exp1}), 
uniform sampling was implemented using the numpy.linspace function, whereas random sampling utilized numpy.random.choice across five different seeds to avoid edge cases.  In some cases, 
%Fig.~\ref{fig:exp2}, 
the number of APDs per measuring point was varied. Because each point may record a different total number of APDs during the measurement period (typically ranging between 900 and 1300), a maximum cutoff of 900 APDs was established. For experiments testing the effects of downsampling, 
%in Fig.~\ref{fig:exp3}, 
the measurement resolution was coarsened to 
%degraded using the formula $dt^{\prime} = 
$M \cdot dt$. Additionally, some experiments 
%Gaussian noise experiment in Fig.~\ref{fig:exp4} 
incorporated Gaussian noise 
%modeled on a normal distribution 
with a mean of 0 and a scale equal to the target noise percentage. Across all experiments, any factor not actively being tuned was held constant at the baseline settings used in prior parameter assessments. For example, when varying the number of APDs per point, the spatial sampling was maintained at the maximum $8\times 8$ grid.

\section{Results}

Below we describe the LE estimate results for the FK and TNNP model for various dynamical regimes. Three different LE calculations are presented: (1) the reference standard results computed from full voltage and gating variable data using the temporal method, $\lambda^T(uvw)$; (2) the APD-based estimate using the same temporal method, $\lambda^T(\text{APD})$; and (3) the APD-based estimation using the spatial-temporal method, $\lambda^S(\text{APD})$. Note that because the spatial-temporal algorithm evaluates one-step divergence, as defined in Eq.~\ref{eq 9}, it is unsuitable for voltage signals that typically exhibit divergence over multiple steps (approximately one APD cycle) and thus is not applied to the full state variable data. Finally, we study the effects of data quality parameters, such as resolution and noise, on the accuracy of LE estimation.

\subsection{Phase Space Reconstruction}

\begin{figure}
    \centering
    {\includegraphics[width=0.46\textwidth]{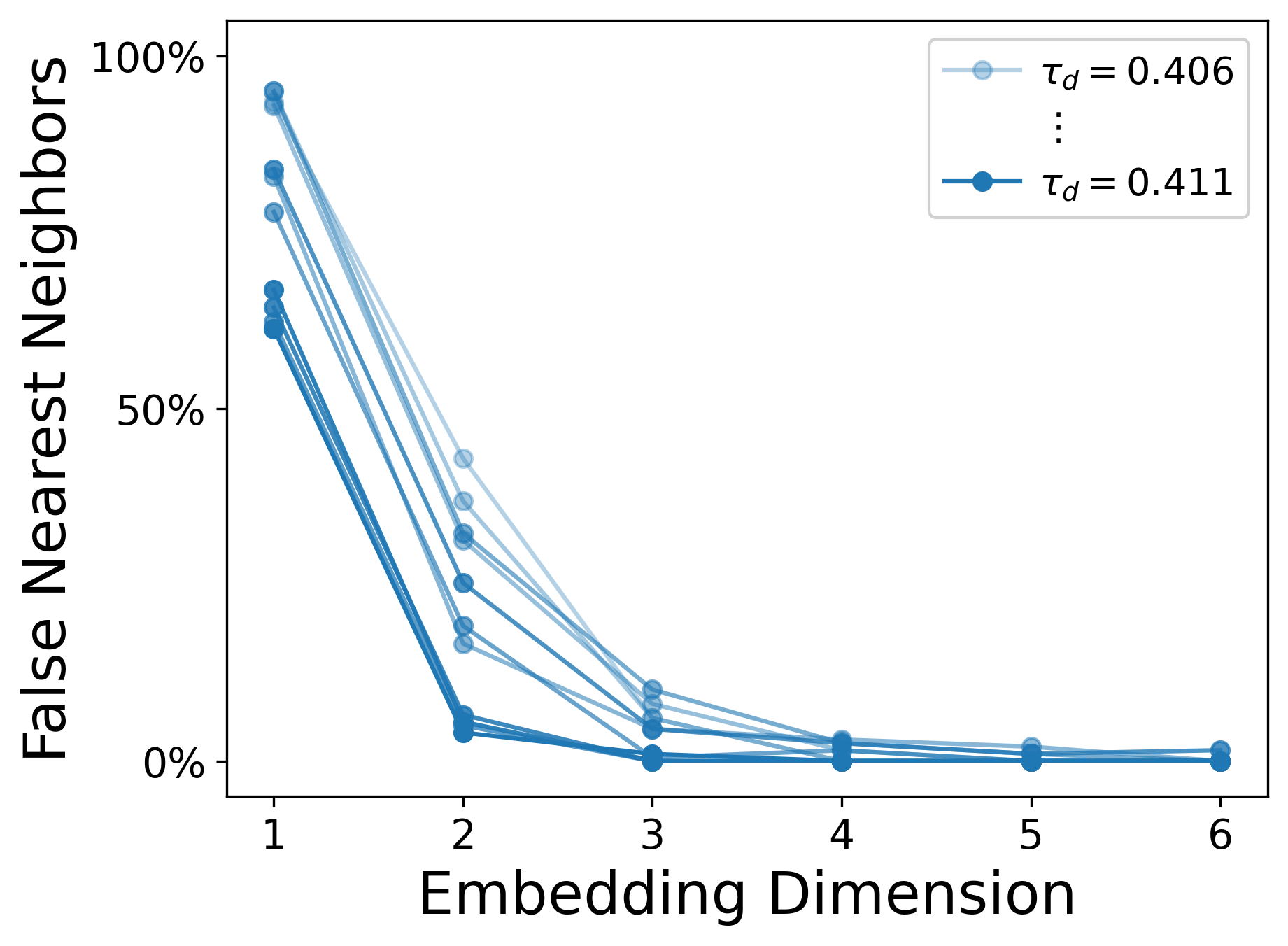}} 
    \caption{Determining embedding dimension using the percentage of false nearest neighbors method with input data $X \equiv \text{APD}$ and lag $\tau = 1$. 
    %In this figure, the $\tau_d$ dataset is presented, with other parameter sets exhibiting similar results. 
    Each curve corresponds to a different value of $\tau_d$ in the FK model, from 0.406 (light blue) to 0.411 (dark blue)~ms with a spacing of 0.005~ms (see Table~\ref{Table FK para}). The percentage drops to near zero at $m = 4$, identifying it as the minimum sufficient embedding dimension. Varying other parameters achieved similar results.}
    \label{fig:FNN}
\end{figure}

As noted earlier, a lag value of \( \tau = 1 \) is selected given that the simulation data is noise-free. The effect of the embedding dimension \( m \) on the percentage of false nearest neighbors using APD signals for different values of \( \tau_d \) is shown in Fig.~\ref{fig:FNN}. Convergence is observed beyond $m = 4$, confirming a dimension sufficient for phase space reconstruction. Therefore, we select \( m = 4 \) as the embedding dimension for all subsequent analyses. Consistent convergence at $m=4$ is observed when varying other parameters in both the FK and TNNP models.

\subsection{FK Model}

For the FK model, modifying the gating constants of different ionic currents alters the biophysical behavior of cardiac cells, potentially leading to intercellular instability that amplifies the tissue’s nonlinear response. 
%This cascade can result in increased LE, indicating a transition toward more chaotic dynamics. 
%A key advantage of the FK model is its ability to capture essential electrophysiological behavior using only three state variables per cell, while remaining flexible enough to represent species ranging from mice to humans\cite{2}. This makes it a valuable tool for investigating how ionic perturbations influence the degree of chaos across species. 
In this section, we study how the different LE estimation methods perform in dynamical regimes corresponding to stationary and meandering single spiral waves as well as spiral-wave breakup conditions. 
%Since both single and multiple spiral waves are known to arise during life-threatening arrhythmias such as fibrillation\cite{karma2013physics}, this study begins with characterizing single-spiral wave regimes and evaluates both $\lambda^S$ and $\lambda^T$ to examine their correspondence with different spiral wave patterns.

\subsubsection{Single Spiral Wave Cases}

\begin{figure}
\centering
        \hspace*{0.2cm}
    \includegraphics[width=0.95\linewidth]{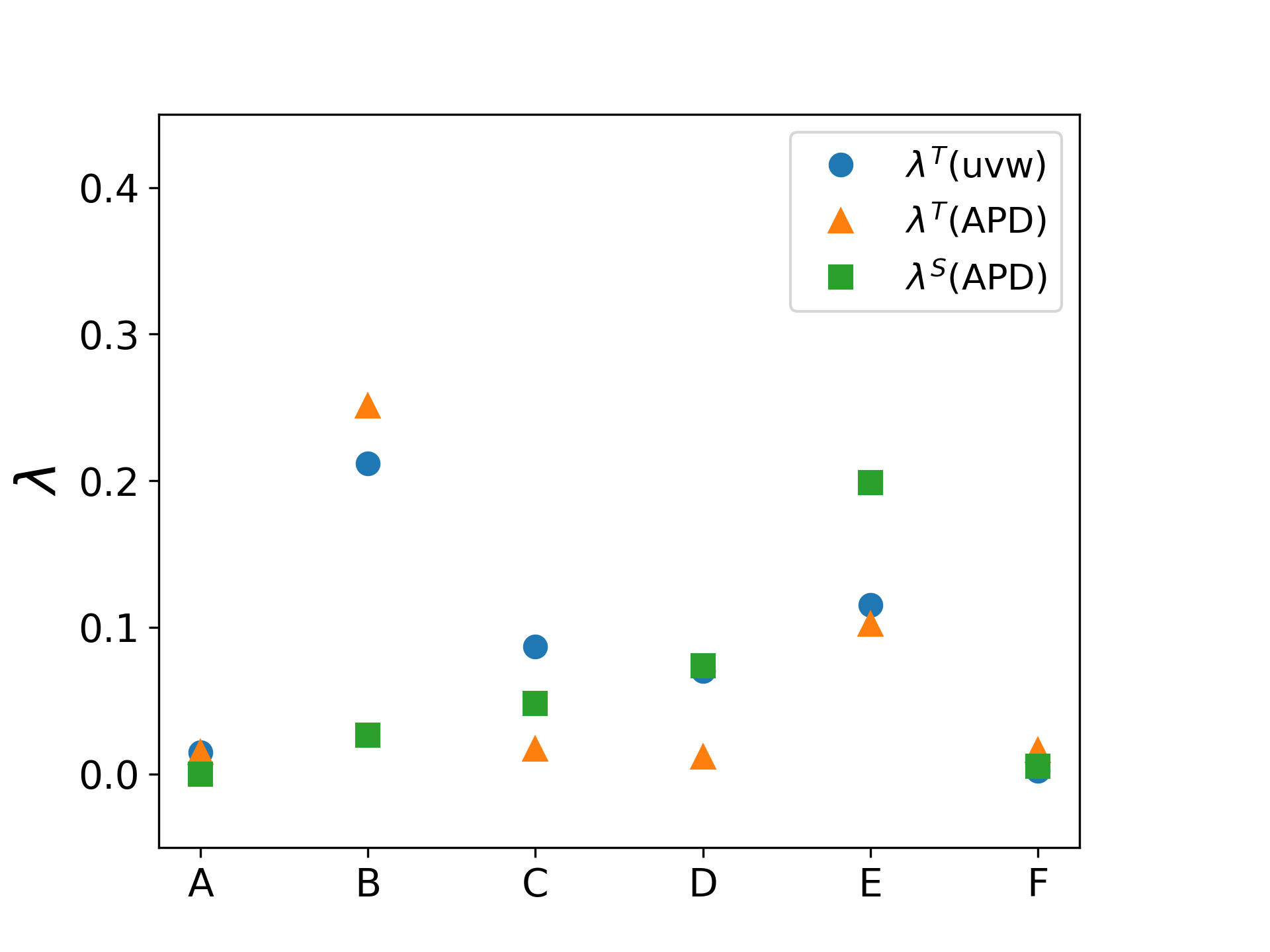}\\
    \subfigure[]{\includegraphics[width=0.14\textwidth]{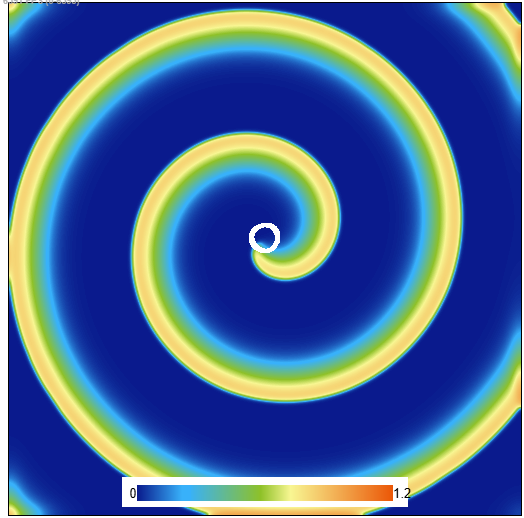}} 
    \subfigure[]{\includegraphics[width=0.14\textwidth]{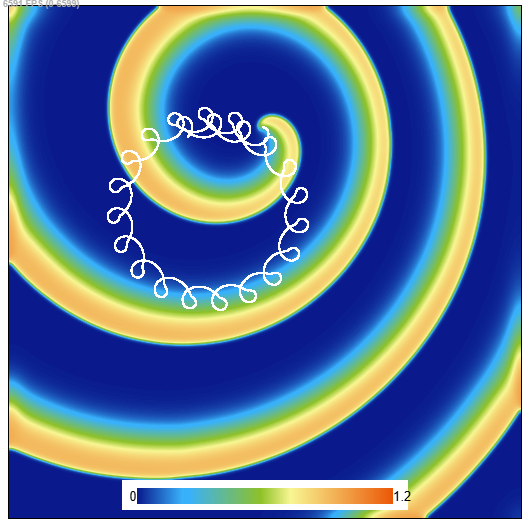}} 
    \subfigure[]{\includegraphics[width=0.14\textwidth]{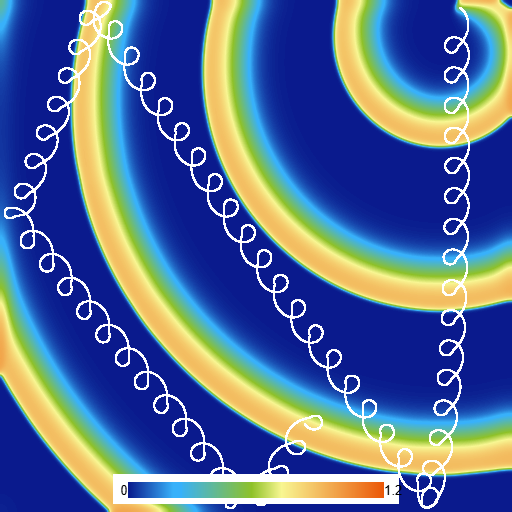}}\\
    \subfigure[]{\includegraphics[width=0.14\textwidth]{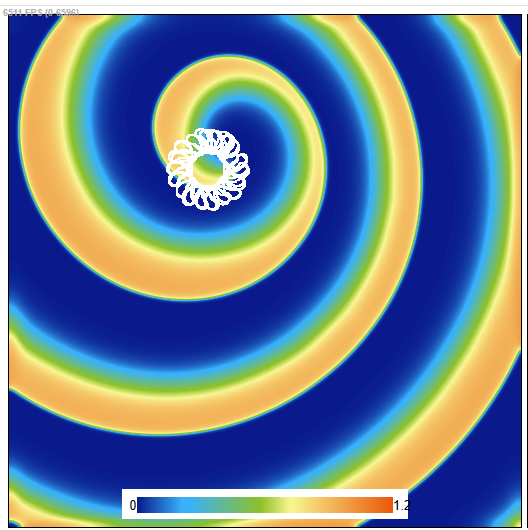}}
    \subfigure[]{\includegraphics[width=0.14\textwidth]{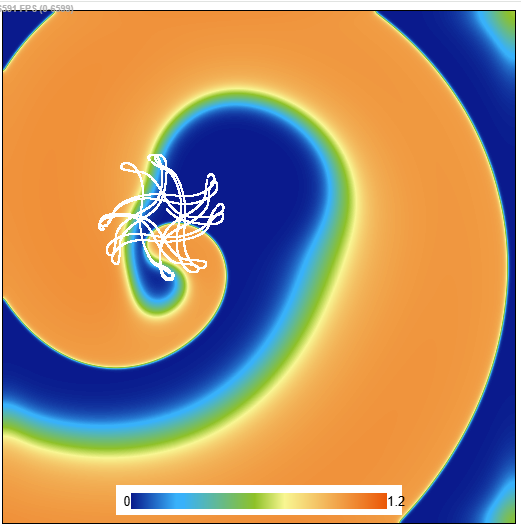}}
    \subfigure[]{\includegraphics[width=0.14\textwidth]{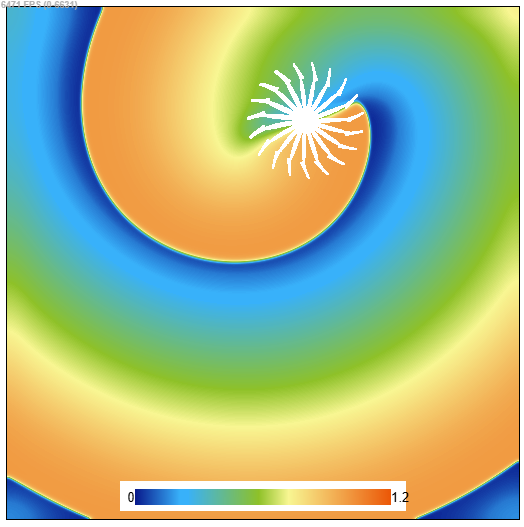}}
\caption{LEs ($\lambda_{\text{single spiral}}$ as shown in Eq.~\ref{Eq TLE_single_spiral} and Eq.~\ref{Eq SLE_single_spiral}) for different single spiral wave dynamics of the FK model. \textbf{Top:} LEs for six different spiral wave dynamics obtained from the state variables ($uvw$, see Eq.~\ref{FK Voltages}) with Wolf’s algorithm ($\lambda^T(uvw)$, blue), 
    from APD signals with Wolf’s algorithm ($\lambda^T(\text{APD})$, orange), and 
    from APD signals with the Spatio-temporal algorithm ($\lambda^S(\text{APD})$, green). \textbf{Bottom:} Simulation snapshots with white curves indicating the spiral core trajectories: 
    (A) circular, 
    (B) epicycloidal, 
    (C) cycloidal, 
    (D) hypocycloidal, 
    (E) hypermeandering, and 
    (F) linear.
    Parameters values are listed in Table~\ref{table: single_spiral_para}. }
    \label{fig: single_spiral}
\end{figure}

\begin{figure}
    \centering
    \subfigure[]{\includegraphics[width=0.15\textwidth]{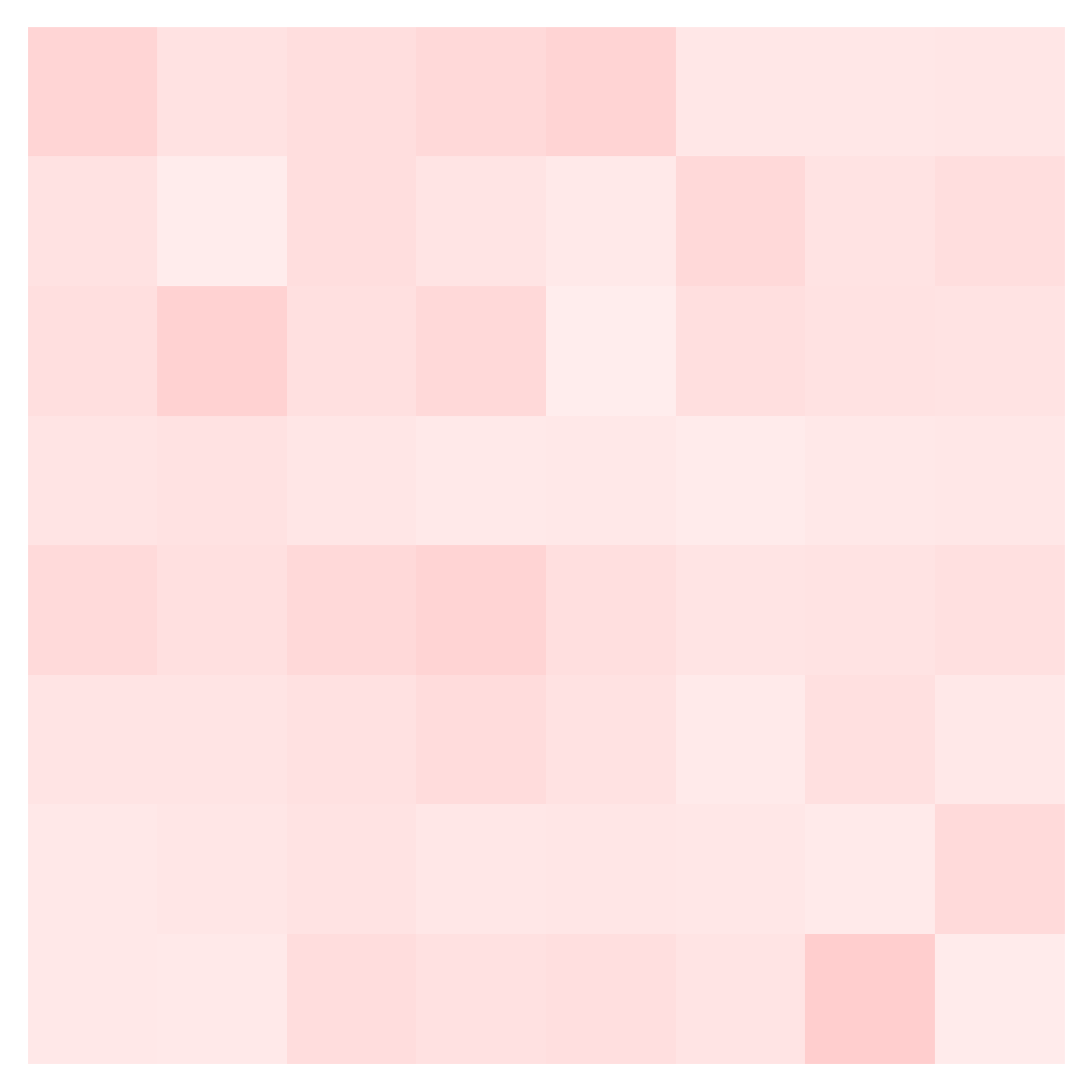}} 
    \subfigure[]{\includegraphics[width=0.15\textwidth]{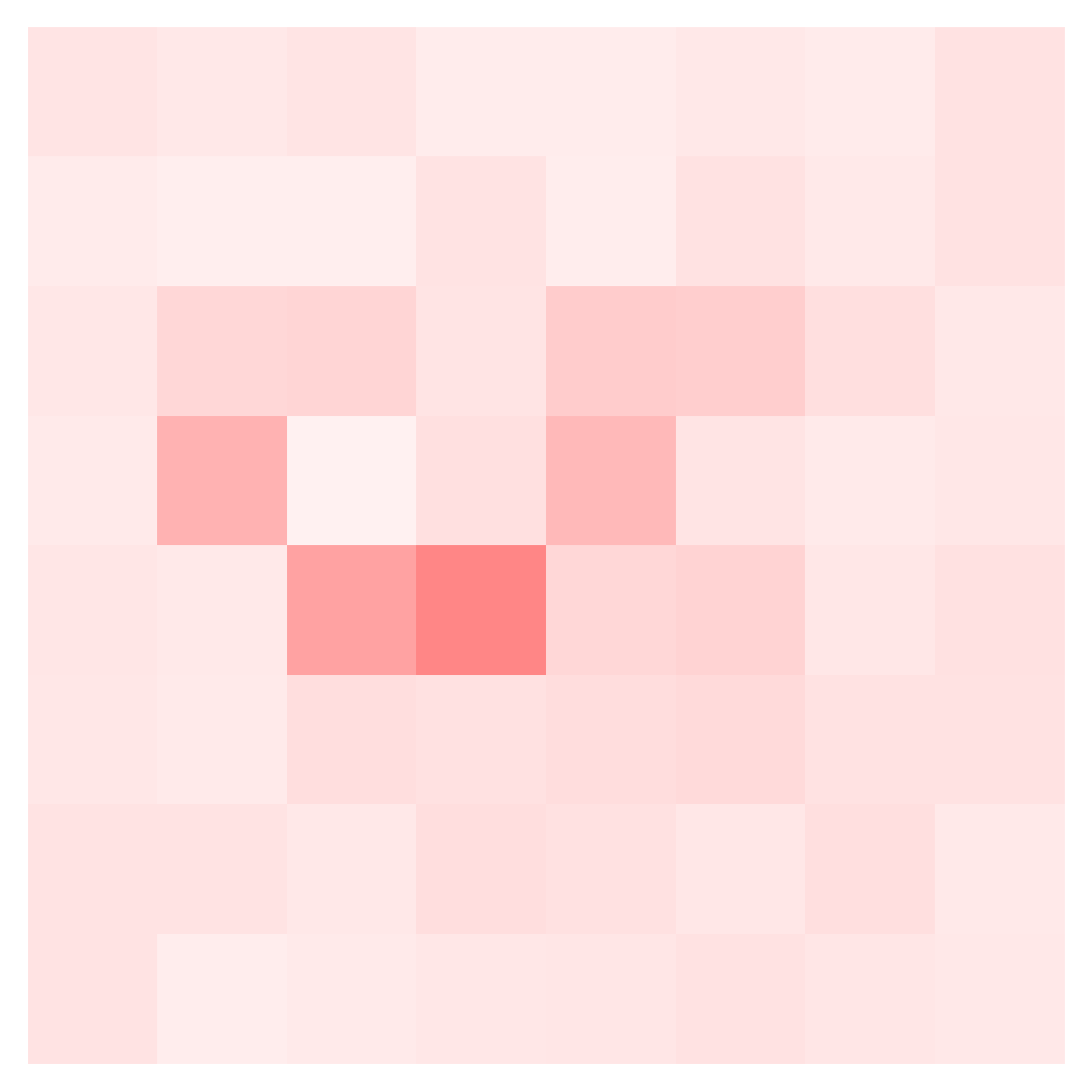}} 
    \subfigure[]{\includegraphics[width=0.15\textwidth]{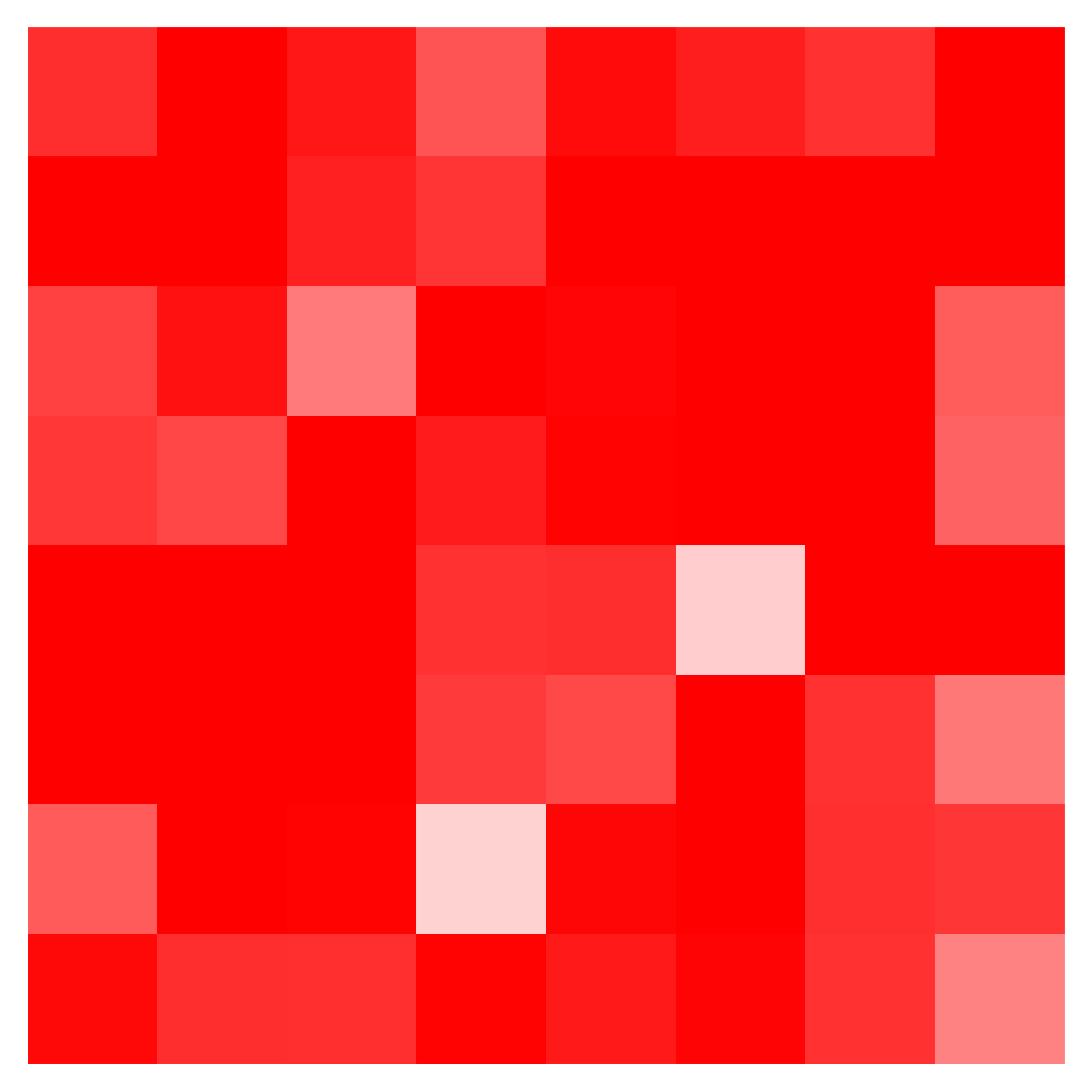}}\\
    \subfigure[]{\includegraphics[width=0.15\textwidth]{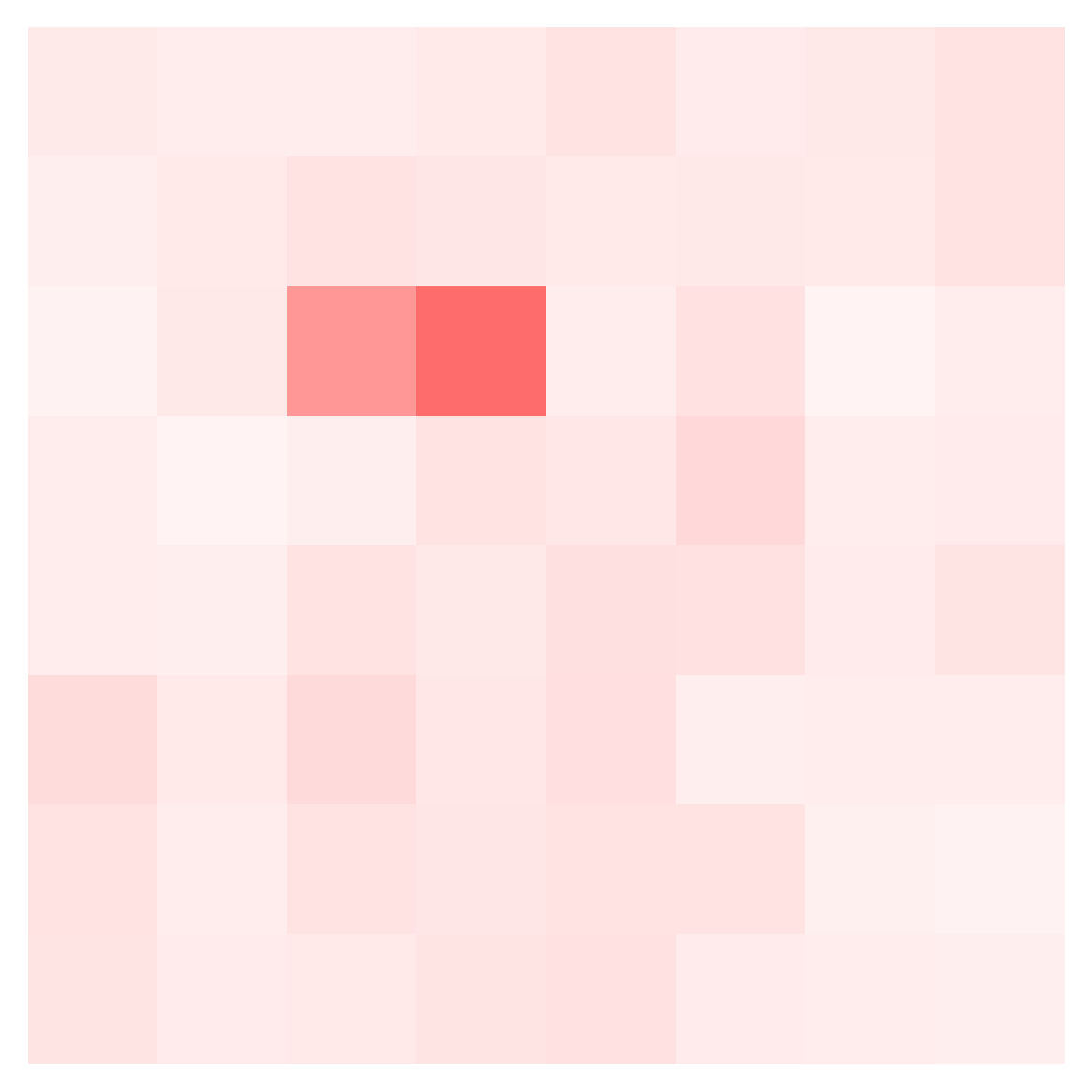}}
    \subfigure[]{\includegraphics[width=0.15\textwidth]{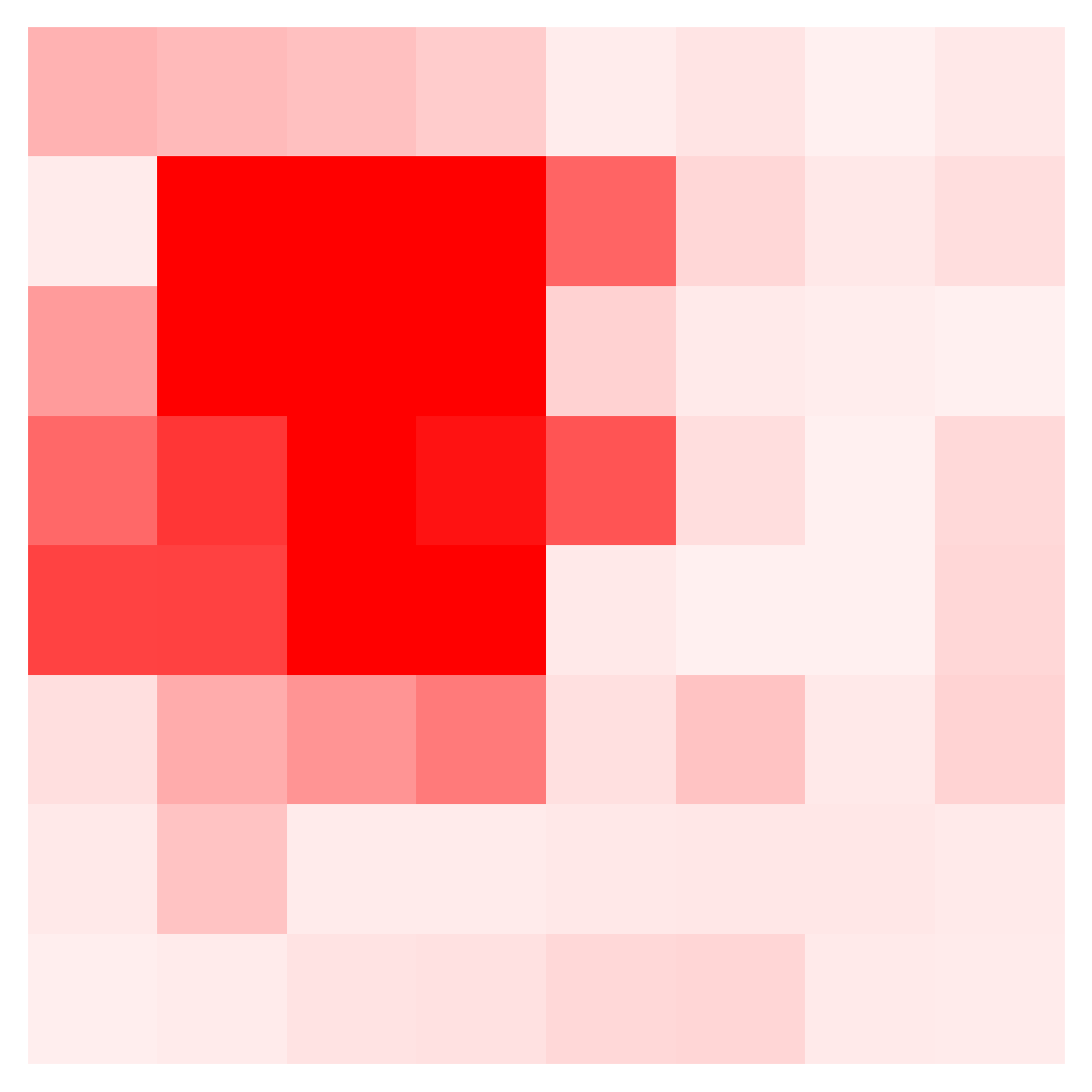}}
    \subfigure[]{\includegraphics[width=0.15\textwidth]{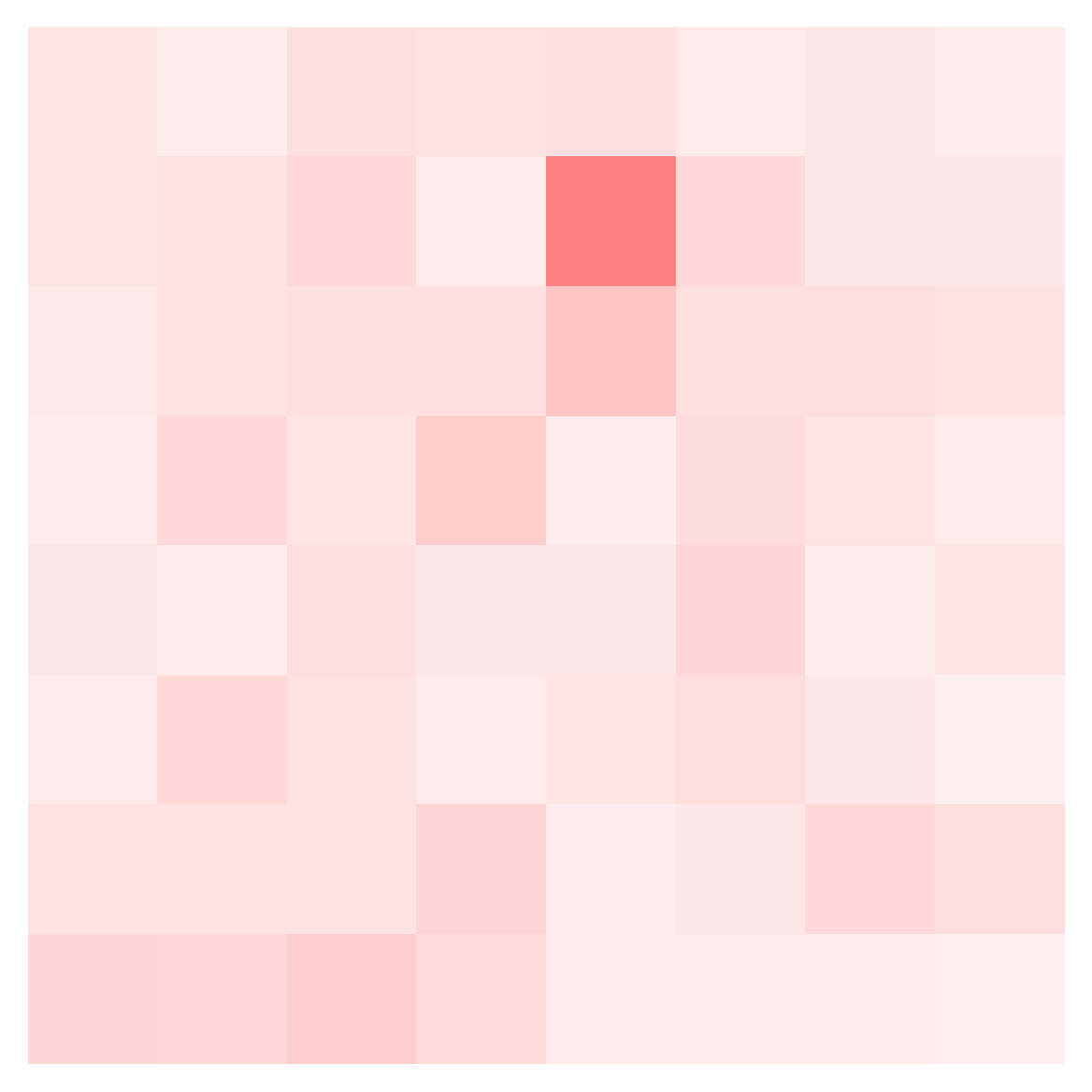}}\\
    \includegraphics[width=0.5\textwidth]{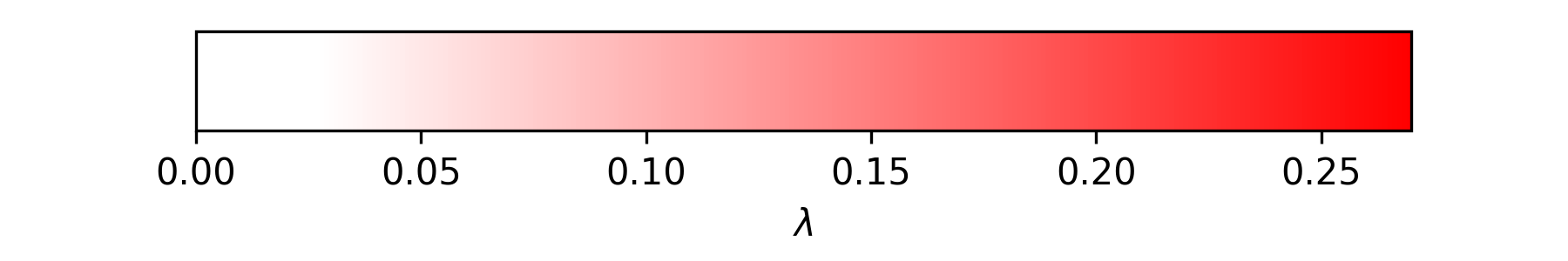}
    \caption{Heatmaps of LEs calculated from APD signals using Wolf’s algorithm ($\lambda^T (\mtx \Gamma(\mtx k))$ at measuring points in Fig.~\ref{fig:initial condition}C, without averaging in Eq.~\ref{Eq TLE_single_spiral}). 
    Panels A–F correspond to spiral wave dynamics shown in Fig.~\ref{fig: single_spiral}.
    Colormap represents local Tempoarl LE value ranging from near zero (blue) to strongly positive (red), highlighting spatial heterogeneity in the degree of chaoticity across the tissue.} %While some dynamics (e.g., circular and linear cores) display relatively uniform distributions, others (e.g., cycloidal and hypermeandering) exhibit localized regions of strongly positive LEs, reflecting their more complex trajectories.}
    \label{fig:heat map}
\end{figure}

\begin{figure}
    \includegraphics[width=0.48\linewidth]{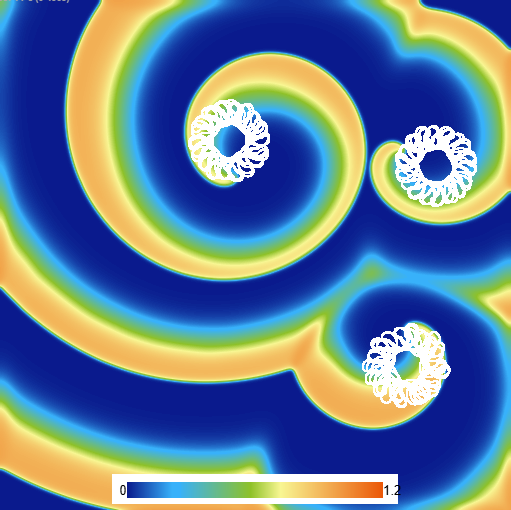}
    \raisebox{-1mm}{\includegraphics[width=0.50\linewidth]{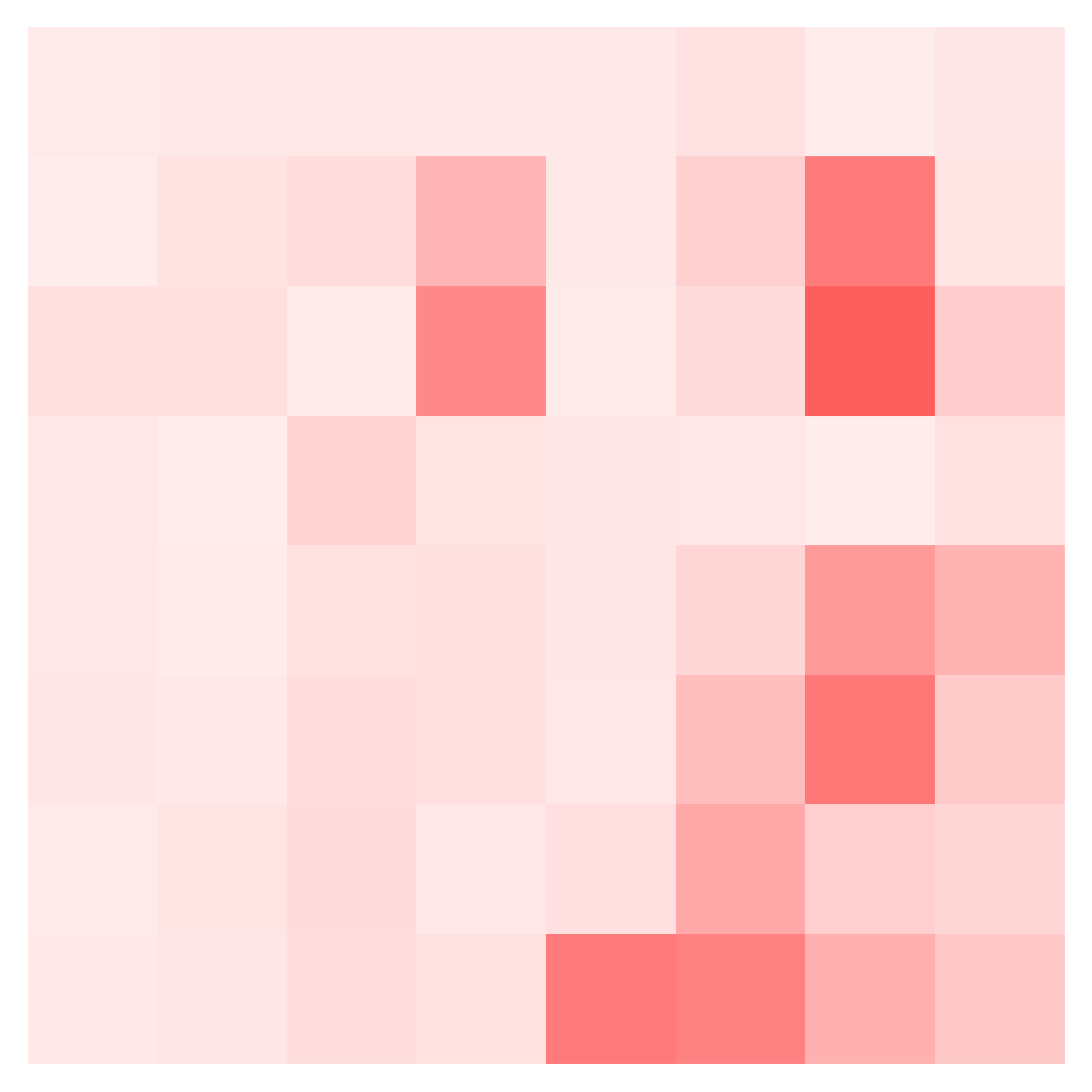}}

    \caption{Example of multi-spiral wave dynamics in case D from Fig.~\ref{fig: single_spiral} with corresponding LE distribution. It was obtained by specifically tailored initial condition that allowed for the formation and sustenance of multiple spiral waves. \textbf{Left:} Simulation snapshot. \textbf{Right:} The corresponding heatmap of LEs (defined as in Fig.~\ref{fig:heat map}), shown with the same colormap for consistency.}
    \label{fig:D2}
\end{figure}

Various types of single spiral wave patterns can be produced by tuning the FK model parameters~\cite{2}, ranging from circular cores (Fig.~\ref{fig: single_spiral}A), to meandering (Fig.~\ref{fig: single_spiral}B--D) and hypermeandering (Fig.~\ref{fig: single_spiral}E) trajectories as excitability is increased, as well as linear cores (Fig.~\ref{fig: single_spiral}F); parameter values are given in Table~\ref{table: single_spiral_para}. These cases display distinct LE estimates depending on the method used, underscoring the need for evaluating both spatial and temporal aspects of instability when characterizing wave dynamics in cardiac tissue.
For example, in the cycloidal trajectory (B), $\lambda^T$ is significantly higher than $\lambda^S$. This discrepancy likely arises from the random walking of the spiral core, as shown in the tip trajectory of Fig.~\ref{fig: single_spiral}B. From a spatial perspective, the overall pattern is a translated version of a periodic spiral wave. However, when observed temporally, without the global view, each cell may be considered to undergo a more irregular process, resulting in a mismatch between $\lambda^S$ and $\lambda^T$.

\begin{table}[htbp]
\centering
\caption{Parameter values of the FK model used to generate the six single spiral wave dynamics shown in Fig.~\ref{fig: single_spiral}. 
%Patterns A–E are obtained from Parameter Set~1 with varying $\tau_d$, while pattern F is produced using Parameter Set~2. 
%The time constants correspond to the following components: $\tau_d$ (fast inward current), $\tau_0,\ \tau_r$ (slow outward current), $\tau_{si}$ (slow inward current), $\tau_v^\pm$ (gating variable $v$), and $\tau_w^\pm$ (gating variable $w$). Additional parameters $k$, $V_c$, $V_v$, and $V_{si}^c$ control activation thresholds and restitution slopes. 
%An ellipsis ($\dots$) indicates that the parameter value is the same as in the preceding column.
All time constants are in units of~ms.
}

\vspace{0.5em}
\begin{tabular}{c|c|c|c|c|c|c}
\hline\hline
\textbf{Parameter} & \textbf{A} & \textbf{B} & \textbf{C} & \textbf{D} & \textbf{E} & \textbf{F} \\
\hline
$\tau_d$           & 0.41 & 0.381  & 0.389  & 0.36   & 0.25   & 0.25 \\ \hline
%$\tau_v^+$         & 3.33 & \ldots & \ldots & \ldots & \ldots & 10 \\
$\tau_v^+$         & \multicolumn{5}{c|}{3.33} & 10 \\ \hline
%$\tau_{v1}^-$      & 19.6 & \ldots & \ldots & \ldots & \ldots & 10 \\
$\tau_{v1}^-$      &  \multicolumn{5}{c|}{19.6}  & 10 \\ \hline
%$\tau_{v2}^-$      & 1000 & \ldots & \ldots & \ldots & \ldots & 10 \\
$\tau_{v2}^-$      & \multicolumn{5}{c|}{1000} & 10 \\ \hline
%$\tau_w^+$         & 667  & \ldots & \ldots & \ldots & \ldots & \ldots \\
$\tau_w^+$         & \multicolumn{5}{c|}{667} & 667 \\ \hline
%$\tau_w^-$         & 11   & \ldots & \ldots & \ldots & \ldots & \ldots \\
$\tau_w^-$         & \multicolumn{5}{c|}{11}  & 11 \\ \hline
%$\tau_0$           & 8.3  & \ldots & \ldots & \ldots & \ldots & 10 \\
$\tau_0$           & \multicolumn{5}{c|}{8.3} & 10 \\ \hline
%$\tau_r$           & 50   & \ldots & \ldots & \ldots & \ldots & 190 \\
$\tau_r$           & \multicolumn{5}{c|}{50} & 190 \\ \hline
%$\tau_{si}$        & 45   & \ldots & \ldots & \ldots & \ldots & \ldots \\
$\tau_{si}$        & \multicolumn{5}{c|}{45} & 45 \\ \hline
%$k$                & 10   & \ldots & \ldots & \ldots & \ldots & \ldots \\
$k$                & \multicolumn{5}{c|}{10} & 10 \\ \hline
%$V_c^{si}$         & 0.85 & \ldots & \ldots & \ldots & \ldots & \ldots \\
$u_c^{si}$         & \multicolumn{5}{c|}{0.85} & 0.85 \\ \hline
%$V_c$              & 0.13 & \ldots & \ldots & \ldots & \ldots & 0.13 \\
$u_c$              & \multicolumn{5}{c|}{0.13} & 0.13 \\ \hline
%$V_v$              & 0.055& \ldots & \ldots & \ldots & \ldots & \ldots \\
$u_v$              & \multicolumn{5}{c|}{0.055} & 0.055 \\
\hline\hline
\end{tabular}
\label{table: single_spiral_para}
\end{table}

In contrast, in cases A and F, both the spatial and temporal LEs are nearly identical and close to zero, indicating globally periodic behavior. In the linear core case (F), although the tip trajectory is more complex than the circular core scenario, it nevertheless  repeats to produce a regular spatiotemporal pattern. Similarly, in patterns B, D and E, both LE measures are in closer agreement than for the cylcoidal case. Small discrepancies still appear because the spiral cores have larger radius compared to the circular case (A) 
%and B, 
and their structures are more complex than the linear case (F). The increased complexity in their core shapes leads to slightly more chaotic behavior in $\lambda^S$. Meanwhile, since only a limited area near the core experiences irregularity, the average $\lambda^T$ remains smaller than $\lambda^S$.

This difference in these two methods is further supported by the $\lambda^T$ heatmaps in Fig.~\ref{fig:heat map}. For circular (A) and linear (F) trajectories, the maps are primarily blue, indicating low instability. For the cycloidal trajectory (B), In pattern B, higher LE values are distributed more broadly, corresponding to the tip’s displacement across the domain. In patterns C, D, and E, the higher LE values are localized around the spiral cores, reflecting the fact that only cells near the tip experience high temporal divergence. Notably, for the hypermeandering trajectory in panel E, a pronounced high-LE region appears near the core, surrounded by low-LE regions elsewhere, highlighting the spatial localization of chaos captured by the $\lambda^T$.

These results confirm that the two methods capture distinct aspects of spiral dynamics: $\lambda^T$ highlights core displacement and local time-wise irregularity, while $\lambda^S$ emphasizes geometric complexity in the wave structure. When the core position and shape are stable (A, F), both estimators converge toward zero. Instability in either aspect produces differences between the methods, providing insight into the underlying source of chaos. If both position and shape are unstable, the estimators instead show consistent positive values, indicating spatiotemporal chaos.

In addition to the single-spiral dynamics, the hypocycloidal trajectory scenario can also support multi-spiral activity under a different initial condition, as illustrated in Fig.~\ref{fig:D2}. 
Unlike the relatively localized instability observed in the single-spiral case, this configuration develops several interacting cores whose breakups sustain widespread disordered activity across the tissue. 
The LEs for this special case are respectively 0.16, 0.031, 0.11 for results from $\lambda^T(uvw)$, $\lambda^T(\text{APD})$ and $\lambda^S(\text{APD})$. The $\lambda^T(\text{uvw})$ and $\lambda^S(\text{APD})$ values are significantly higher than those in case D (where both are $< 0.1$), reflecting a stronger spatiotemporal chaos induced by the coexistence of multiple spirals. 
This special case highlights how even within a parameter regime classified as single-spiral dynamics, the system may spontaneously transition into a multi-spiral state, thereby bridging our analysis of isolated spirals with the more complex patterns considered next.

\subsubsection{Multiple Spiral Wave Cases}

The transition from single to multiple spiral waves can arise from varying key parameters individually.
%occurs when certain identifiable parameters vary within specific ranges. 
To study the changes in LE values across these transitions, we separately vary the time constants 
$\tau_d$, $\tau_0$, $\tau_r$, and $\tau_{si}$ 
over the ranges specified in Table~\ref{Table FK para}. (results for $\tau_0$ and $\tau_{si}$ are discussed in Appendix A).

\begin{table}[htbp]
\centering
\caption{Parameter values of the FK model used to study transitions in spiral wave dynamics. Along with their base values, the time constants $\tau_d$, $\tau_0$, $\tau_r$, and $\tau_{si}$ also include the ranges studied in Sec.~IV.B.2. Time constant units are~ms.}
%to probe the transition from periodic state to chaos. All other parameters were held fixed at the values of Set~4 from Fenton et al.~\cite{2}.}
\label{Table FK para}
\renewcommand{\arraystretch}{1.2}
\begin{tabular}{lcc}
\hline\hline
\textbf{Parameter} & \textbf{Value} & \textbf{Range} \\
\hline
$\tau_v^+$ & 3.3 & -- \\ 
$\tau_{v1}$ & 16 & -- \\ 
$\tau_{v2}$ & 5 & -- \\ 
$\tau^+_{w}$ & 350 & -- \\ 
$\tau^-_{w}$ & 80 & -- \\ 
{$\tau_{d}$} & 0.41 & [0.406, 0.411] \\ 
{$\tau_{0}$} & 9 & [7.2, 9.4] \\ 
{$\tau_{r}$} & 34 & [33.5, 34.5] \\ 
{$\tau_{si}$} & 27 & [26.45, 26.95] \\ 
$k$ & 15 & -- \\ 
$u^{si}_c$ & 0.45 & -- \\ 
$u_c$ & 0.15 & -- \\ 
$u_v$ & 0.04 & -- \\
\hline\hline
\end{tabular}
\end{table}

\begin{figure}
    \centering
    {\includegraphics[width=0.48\textwidth]{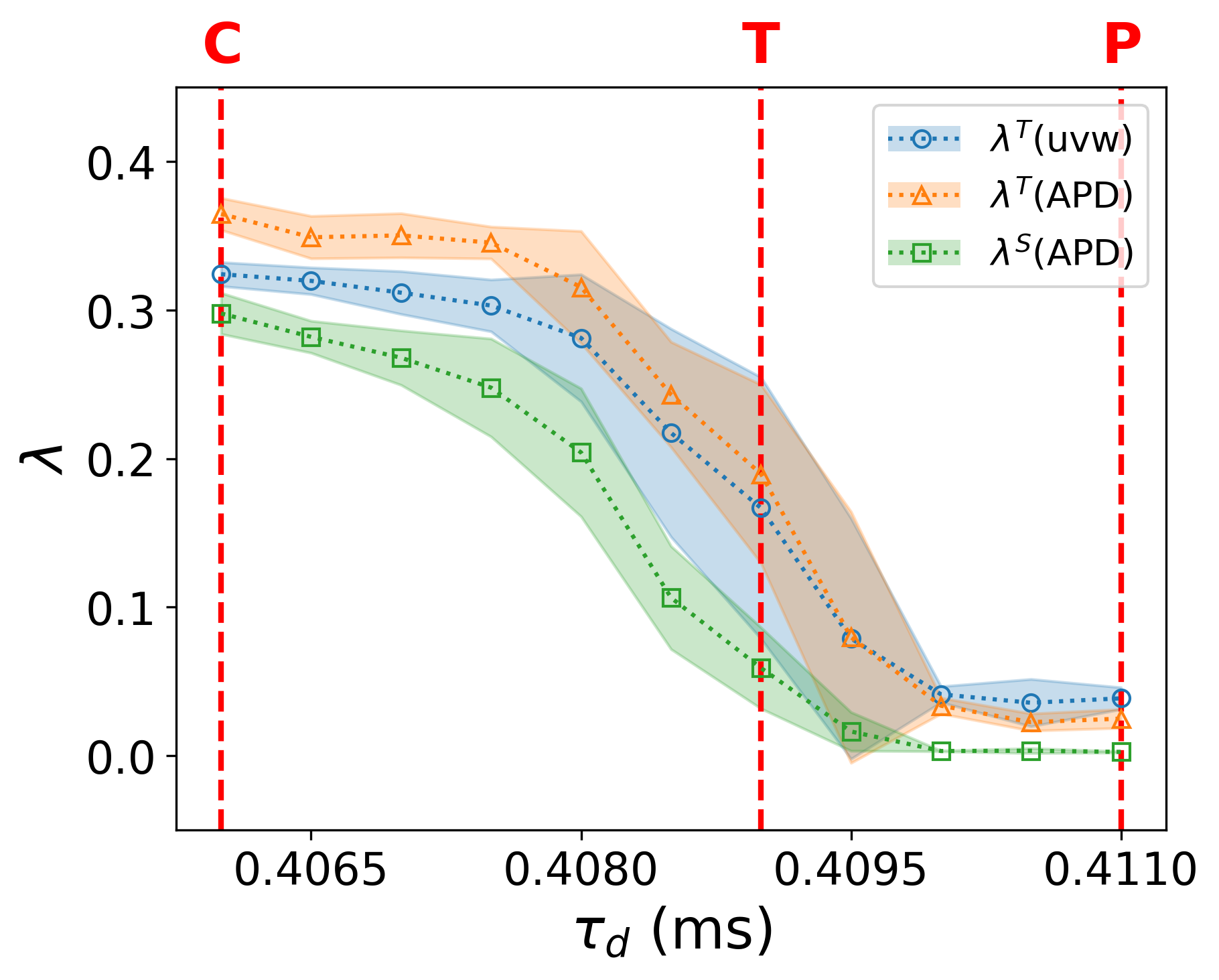}} 
    {\includegraphics[width=0.5\textwidth]{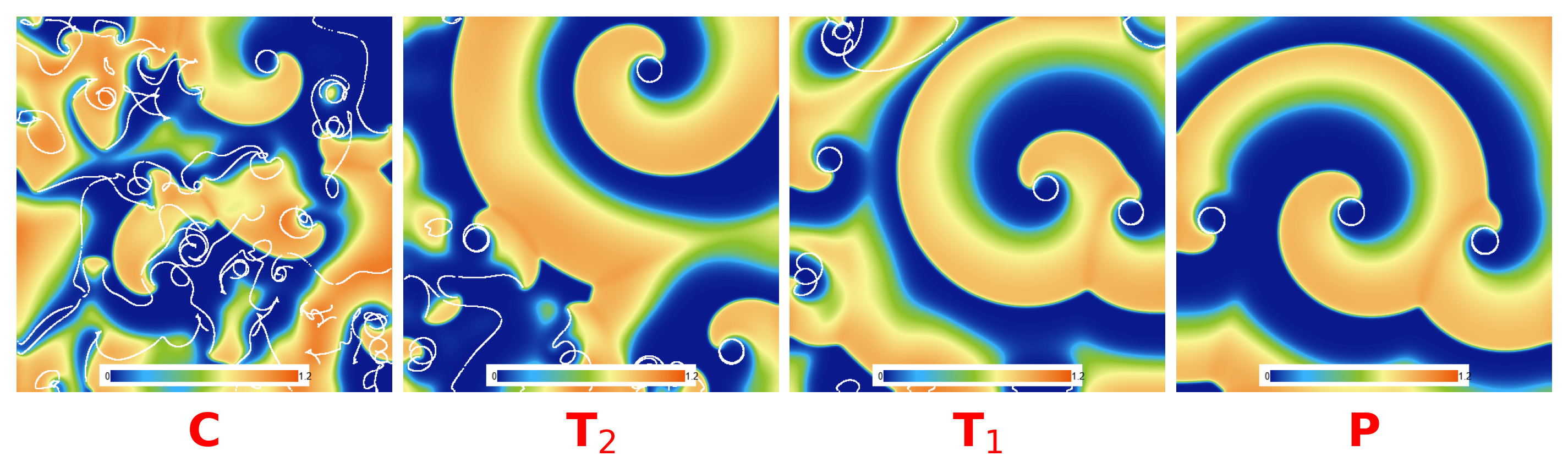}}
    {\includegraphics[width=0.45\textwidth]{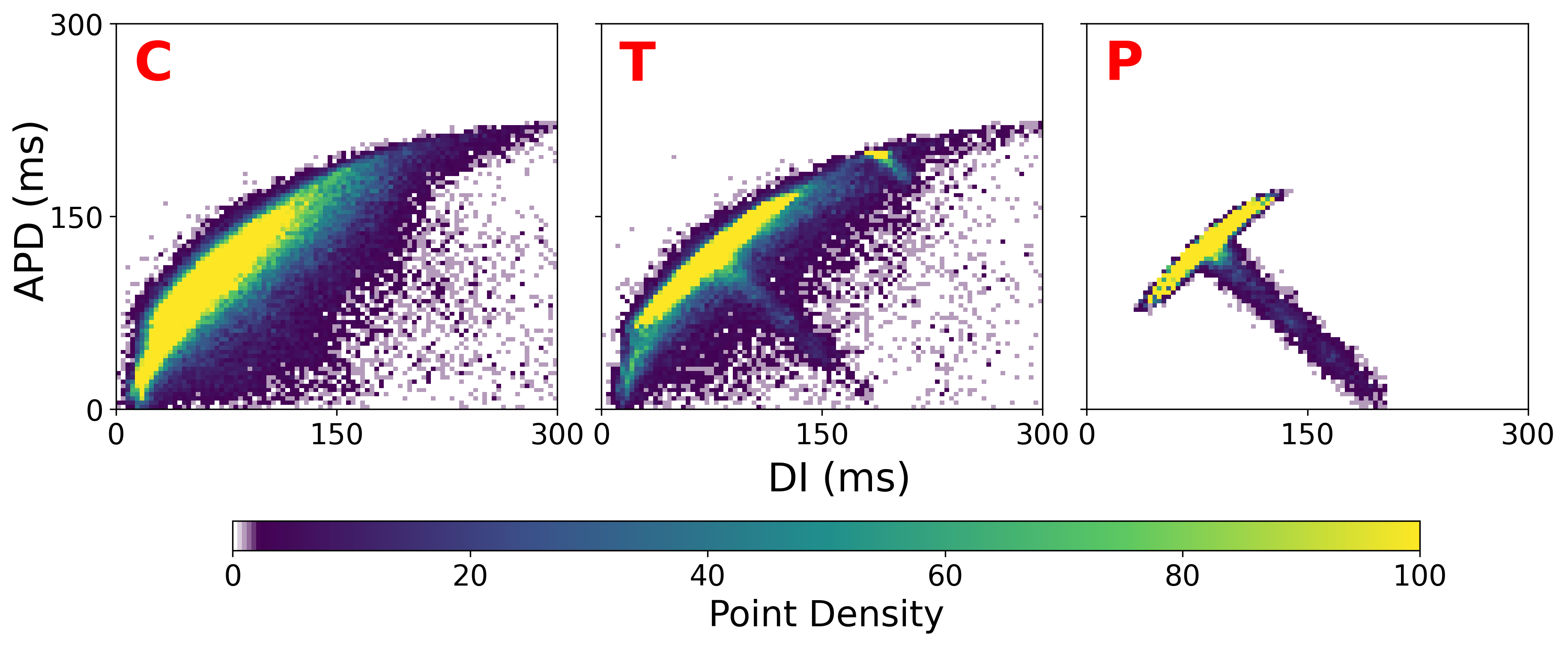}} 
    \caption{Performance of LE estimates for different spiral wave dynamics obtained by varying $\tau_d$ in the FK model. 
    \textbf{Top:} LEs calculated from the full state variables ($u,v,w$) and from APDs only. The shaded bands indicate the standard error derived from averaging $\lambda^T$ and $\lambda^S$ across the across five temporal zones, which are $z = 1$ to $z=5$ in Eq.~\ref{Eq TLE_multi_spiral} and Eq.~\ref{Eq SLE_multi_spiral}. 
    %$\lambda^T(\text{APD})$ and $\lambda^T(uvw)$ are obtained using Wolf’s algorithm, while $\lambda^S(\text{APD})$ are from the spatio-temporal method. 
    Labeled points correspond to example values of $\tau_d$ within the chaotic (C), transitional (T), and periodic (P) regimes. 
    \textbf{Middle:} Representative simulation snapshots for the three values of $\tau_d$ indicated in the top plot with the corresponding spiral tip trajectories shown in white. For the transitional regime, two cases can be observed during one simulation: chaotic breakup (T$_2$) and a moderate one with less breakup (T$_1$). 
    \textbf{Bottom:} Restitution curves of APD versus DI for the three regimes. The chaotic state (A) shows broad APD dispersion, the transitional state (B) exhibits mixed quasiperiodic/chaotic features, and the periodic state (C) displays a narrow and stable APD distribution.}

    \label{fig:taud}
\end{figure}

In Figure~\ref{fig:taud}, we investigate the effects of varying the excitability $\tau_d$.
%controls the capacity of the gate of $I_{fi}$ or $Na^+$ current, as shown in Eq.~\ref{FK Current}. 
We present the results for varying $\tau_d$ from 0.406 to 0.411, comparing APD-based LEs ($\lambda^T(\text{APD})$ and $\lambda^S(\text{APD})$ with standard $uvw$ results ($\lambda^T(uvw)$). We identify three dynamic regimes—chaotic (C), transitional (T), and periodic (P)—to characterize the system’s chaosity as model parameters are varied. C represents the state where the LE reaches its highest plateau and the simulation shows the most spiral wave breakups. The representative point, as indicated by the red dashed line in Fig.~\ref{fig:taud} for C, is selected where the LE is highest. P is identified where the LE reaches its lowest plateau, approaching zero, which corresponds to repeating spiral patterns with virtually no wave breakups. Its representative point is chosen when the LE is at its minimum. T is defined as the region between C and P. The representative point for T is selected from the midpoint of this transition region. In the transition region, mixed behavior can be observed during one simulation, including chaotic state with frequent breakups (T$_{2}$) and moderately chaotic state with significantly fewer breakups (T$_{1}$). These observations collectively demonstrate a clear transition from chaos to periodicity. Notably, the APD-based LE results closely align with those derived from full-state data, confirming that APD preserves information about the system's underlying dynamics.

\begin{figure}
    \centering
    {\includegraphics[width=0.48\textwidth]{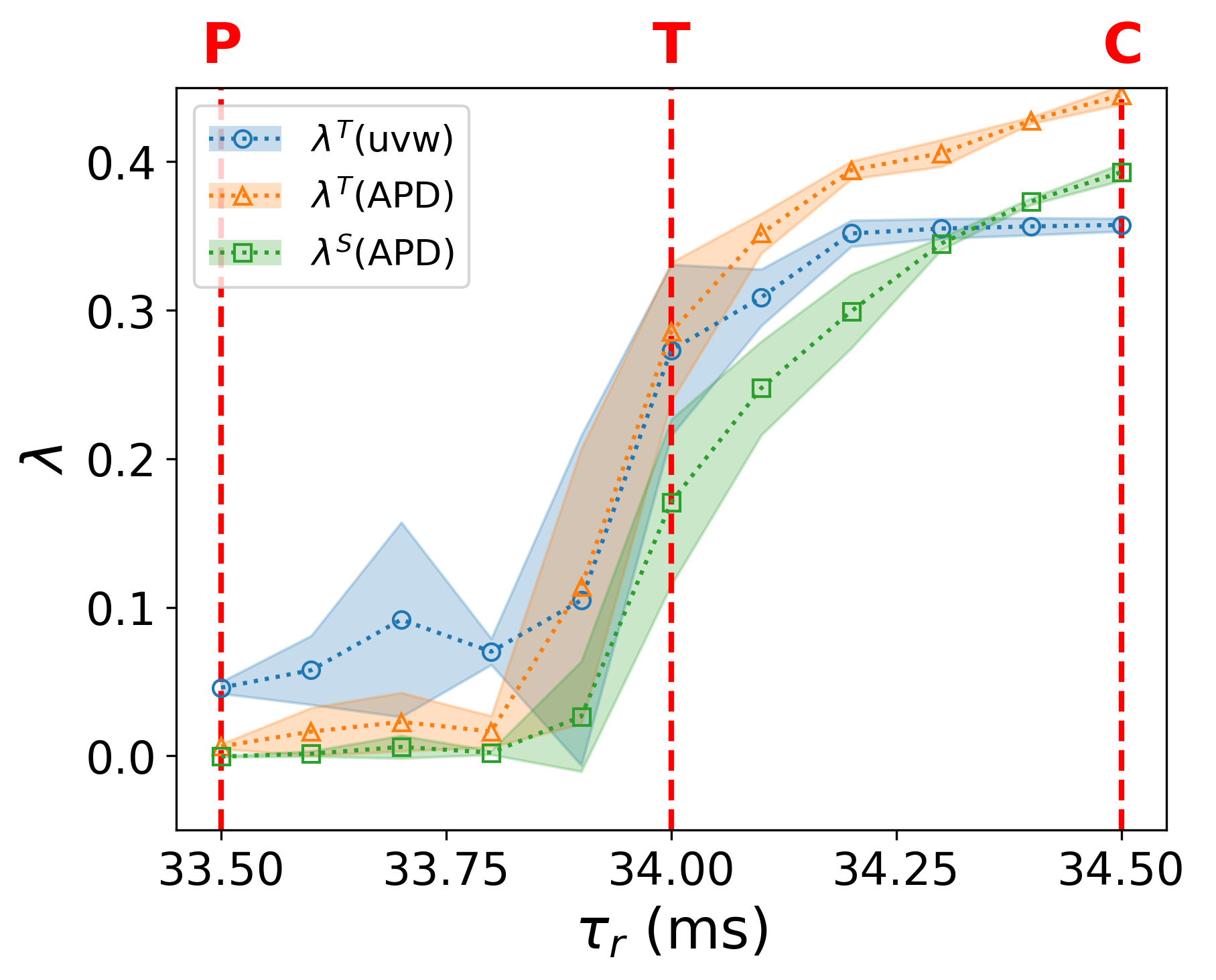}} \\
    {\includegraphics[width=0.5\textwidth]{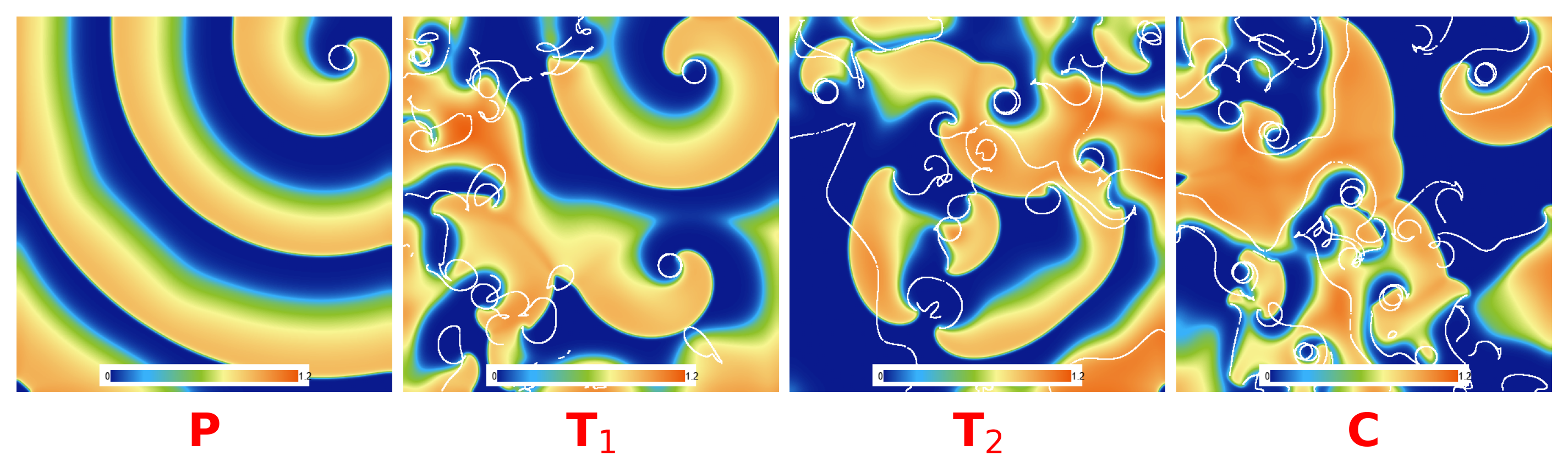}} \\
    {\includegraphics[width=0.45\textwidth]{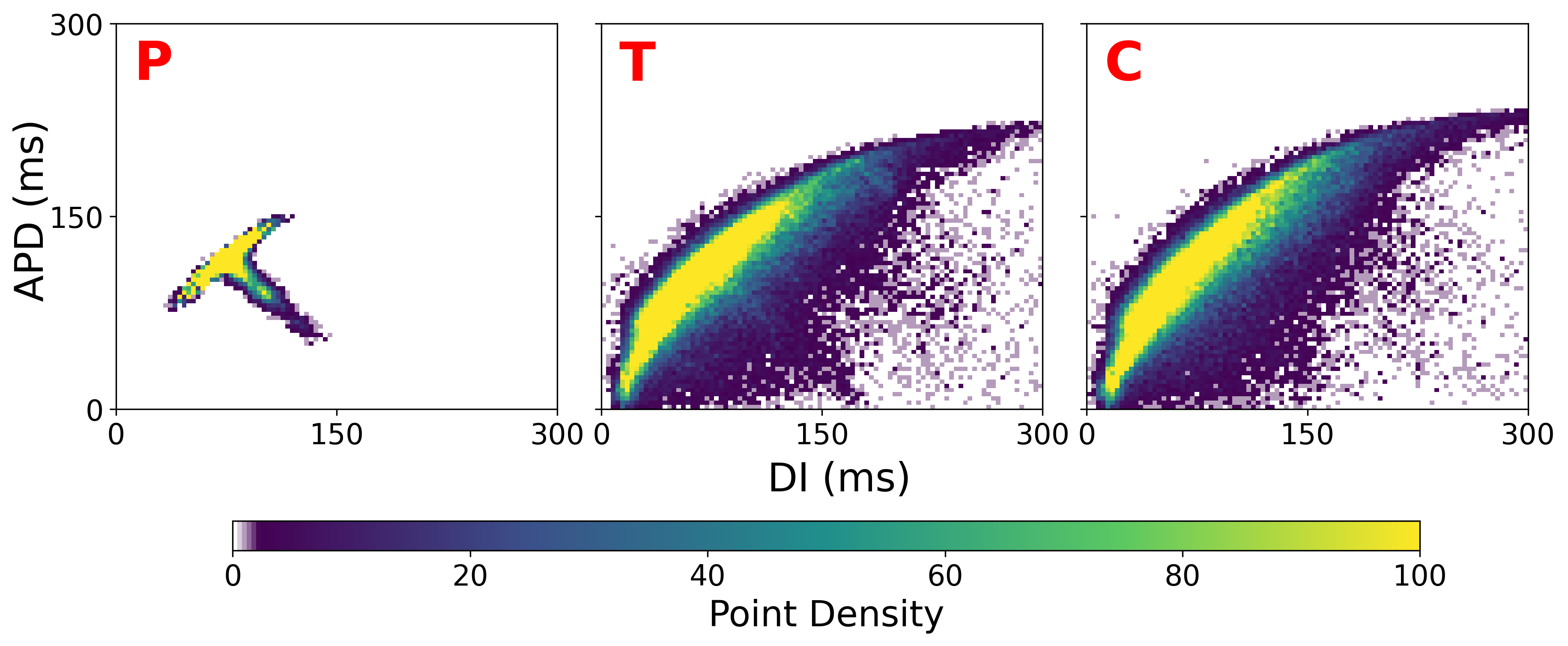}} 
\caption{Performance of LE estimates for different spiral wave dynamics obtained by varying $\tau_r$ in the FK model. 
\textbf{Top:} LEs from state variables and APD signals, as in Fig.~\ref{fig:taud}. 
Labeled points correspond to example values of $\tau_r$ within the periodic (P), transitional (T), and chaotic (C) regimes. 
\textbf{Middle:} Simulation snapshots with spiral tip trajectories (white). 
\textbf{Bottom:} Restitution curves of APD versus DI for the different states, plotted with range [0,300] on both axes. The broader APD spread in regime T reflects the enhanced chaoticity captured in the LE results.}
    \label{fig:taur}
\end{figure}

Figure~\ref{fig:taur} presents the results for varying $\tau_r$. T exhibits behavior similar to that observed for the $\tau_d$ case. For larger $\tau_r$ values, the APD-based LEs reach up to $\lambda \approx 0.45$, higher than the values observed in the $\tau_d$ case and later shown for $\tau_0$ and $\tau_{si}$ in Appendix A. This higher order of chaos is not reflected in the $uvw$-based method, suggesting that APD signals may capture additional instabilities that are not revealed in the full state-variable analysis, though further work would be needed to confirm whether these correspond to genuine hyperchaotic dynamics.

For both cases, if different dominant attractors with different LEs coexist within a relatively confined range of parameter values, bigger standard deviation in the LE estimates is observed. This is most evident in the transition regimes, where a consistently elevated standard deviation reflects the underlying multistability of the system.

%To assess temporal variability in LE estimates, each post-transient simulation was divided into five equal-duration zoness, and the LE was computed independently for each. The standard deviation of these values is shown as the envelope width of the shaded region, If different dominant attractors with different LE values coexist within a relatively confined range of parameter values, greater variability in the LE estimates is observed. Across all parameter sets, the transition regime consistently exhibits elevated standard deviation, reflecting the system's multistability. 
%This reflects the presence of multiple coexisting attractors, including chaotic, quasiperiodic, and periodic attractors. 
%As a result, even with the same initial conditions and after the transient period, the system can develop different behavior and thus demonstrate a wide range of LEs. This sensitivity further supports the interpretation of the transition regime as a dynamically unstable region.

\subsection{TNNP Model}
To compare our results of the FK model to a more complex ionic cell model, we use the TNNP model~\cite{ten2006alternans}. In our simulations, the maximal sodium current conductance ($G_{Na}$) was scaled from 0.5 to 8 while all other parameters remained unchanged. The maximum initial separation, $\epsilon$, was kept identical to that used in the FK model to maintain a consistent scale for the Lyapunov exponents. For the largest reduction in $G_{Na}$, the tissue maintained one or two stable spirals with LEs close to zero. The standard deviation remained quite small for both $\lambda^T(\text{APD})$ and $\lambda^S(\text{APD})$, indicating that a single periodic attractor dominated the dynamics.

\begin{figure}
    \centering
    {\includegraphics[width=0.48\textwidth]{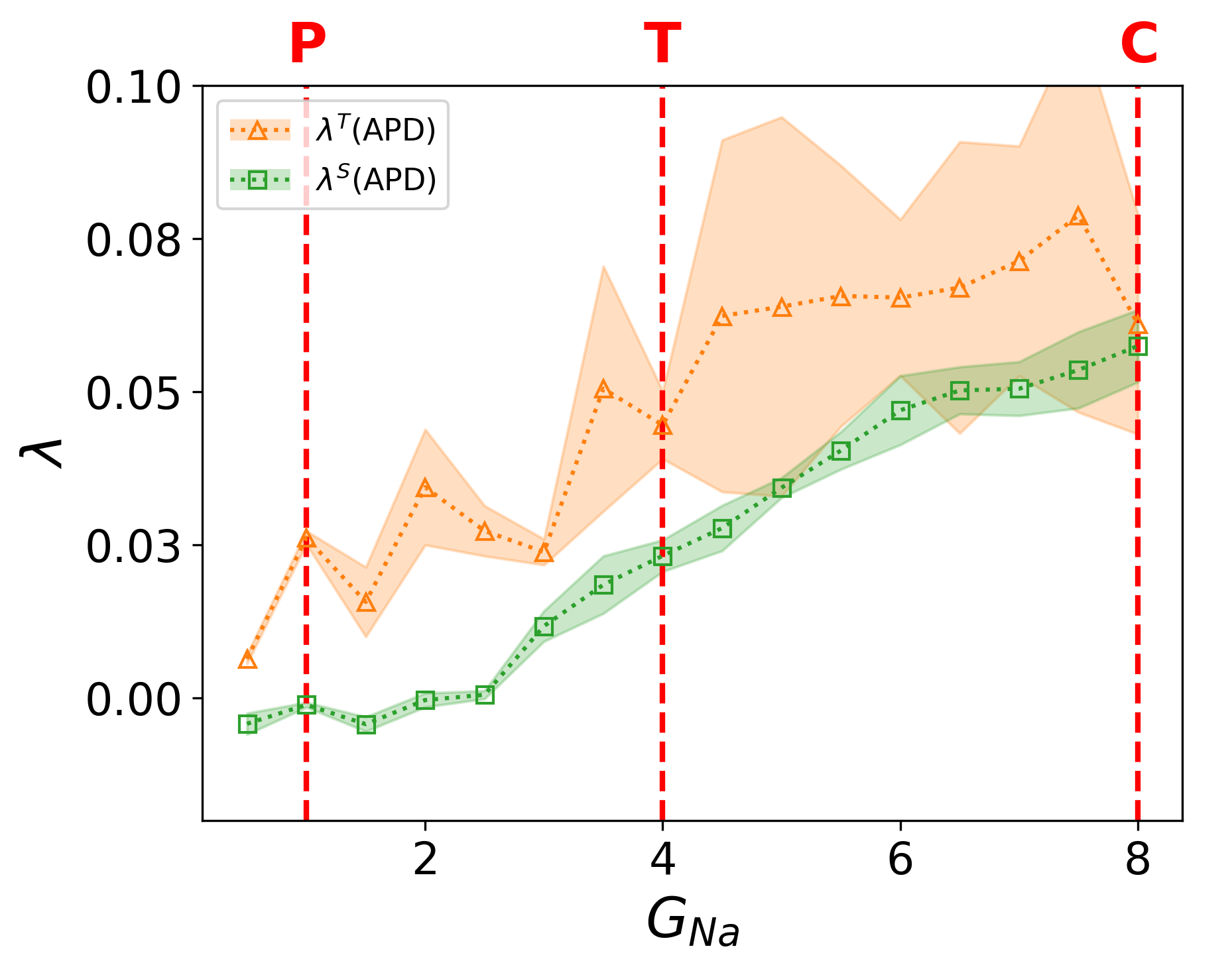}} \\
    {\includegraphics[width=0.5\textwidth]{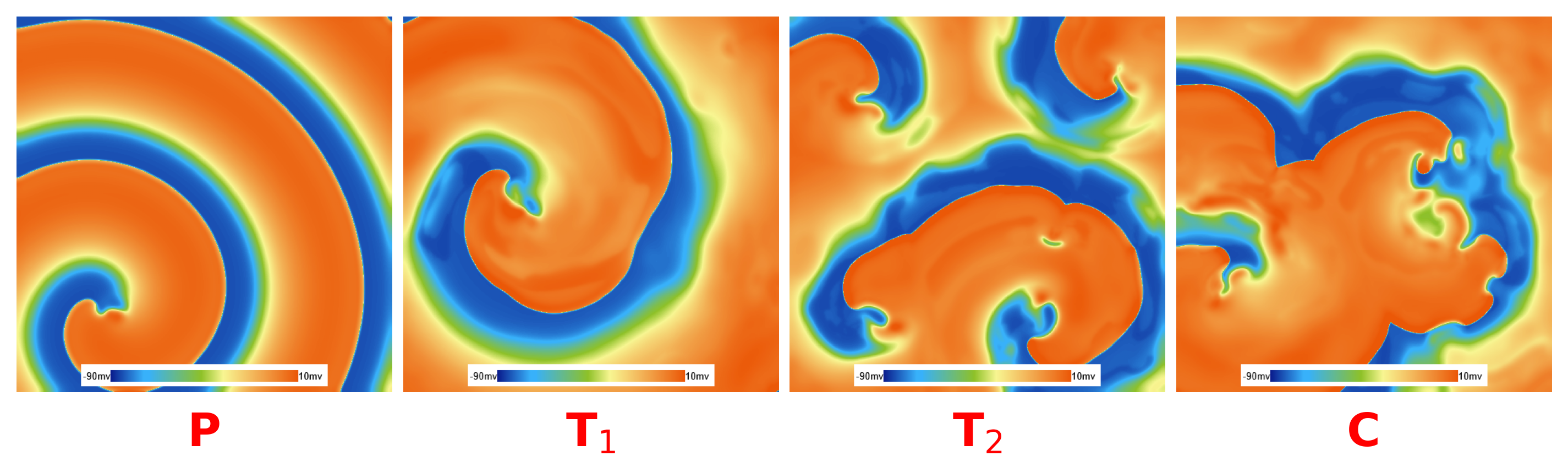}} \\
    {\includegraphics[width=0.45\textwidth]{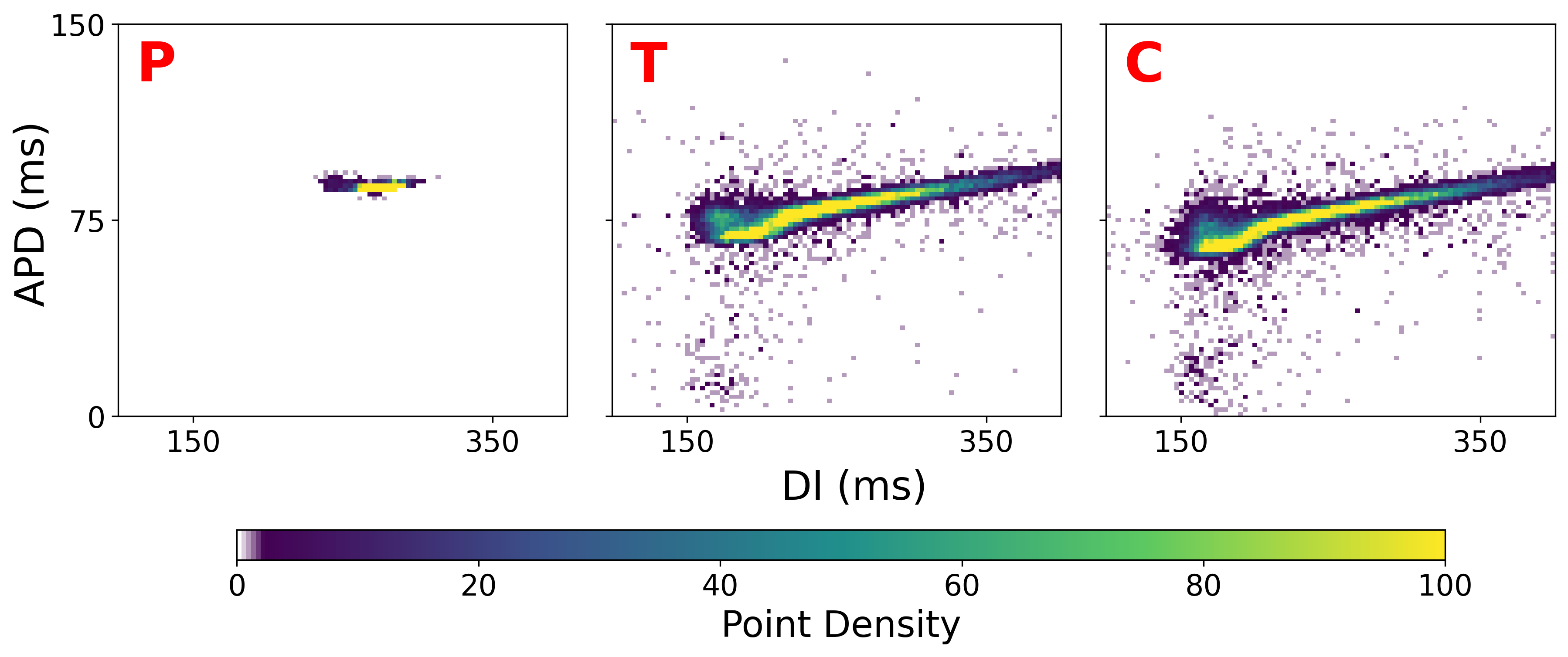}} 
    \caption{Dependence of spiral wave dynamics on the sodium current conductance scaling factor $C_{Na}$ in the TNNP model. To maintain a consistent scale for comparison with the FK model, the maximum initial separation ($\epsilon$) was kept identical.
    \textbf{Top:} LEs from APD signals only. The shaded bands indicate the standard error. %LEs in the FK model range up to 0.4, values for the TNNP model remain below 0.1.
    \textbf{Middle:} Simulation snapshots with spiral tip trajectories (white) illustrating the regimes: (P) periodic, (T) chaotic breakup, and (C) multi-spiral chaos.
    \textbf{Bottom:} Restitution curves of APD versus DI for the identified states.}
     \label{fig:TNNP}
\end{figure}

As $G_{Na}$ was increased, the enhanced conduction velocity and altered restitution properties \cite{ten2006alternans} promoted earlier wavefront collisions and occasional spiral breakup. These changes led to higher LEs, but only $\lambda^T(\text{APD})$ shows a wide range of values measured in different temporal segments (wider envelope), while $\lambda^S(\text{APD})$ is steady across different measurement regions.

For the largest values of $G_{Na}$, the model developed spatiotemporal chaos, with frequent spiral-wave breakup (state C in Fig.~\ref{fig:TNNP}). The temporal LE curves displayed substantial variability in time, resulting in a broad shaded region.
%, indicating the model’s greater dimensionality. 
In contrast, the spatial LE measurements showed greater consistency for large $G_{Na}$, 
suggesting their potential for quantifying chaotic states in high-dimensional systems.

However, it is notable that the maximum LE values for both methods in the TNNP model do not exceed 0.1, as opposed to the 0.4 observed in the FK model. This suggests that, under the same LE hyperparameters ($\epsilon$ and $\theta$), the TNNP model scenarios considered exhibit inherently weaker chaotic dynamics, even during breakup. Alternatively, it may indicate that the TNNP model cannot sustain higher levels of chaos long enough for measurement for the parameter values considered, leading to spiral wave self-termination before a reliable LE estimation can be achieved. This suppressed chaotic behavior is also evidenced by the restitution map (Fig.~\ref{fig:TNNP}, bottom), where the APD distribution is largely confined to a narrow interval of 60 to 100~ms relative to the broader distribution of DIs observed.

\subsection{Effect of Data Quality on LE Estimation Accuracy}

Prior studies have emphasized that limited spatial or temporal resolution could potentially obscure important features of spiral wave dynamics, leading to mischaracterization of underlying instabilities~\cite{cherry2008visualization,roney2017spatial,king2017effect,gandara2024effect}. In light of these previous findings, we 
%how observational constraints and stochastic interference to 
examined how data quality affects the accuracy and reliability of LE estimation and aimed to determine the minimal data requirements for robust chaos quantification.

\begin{figure}[htbp!]
    \centering
    \hspace*{-1cm}
    {\includegraphics[width=0.46\textwidth]{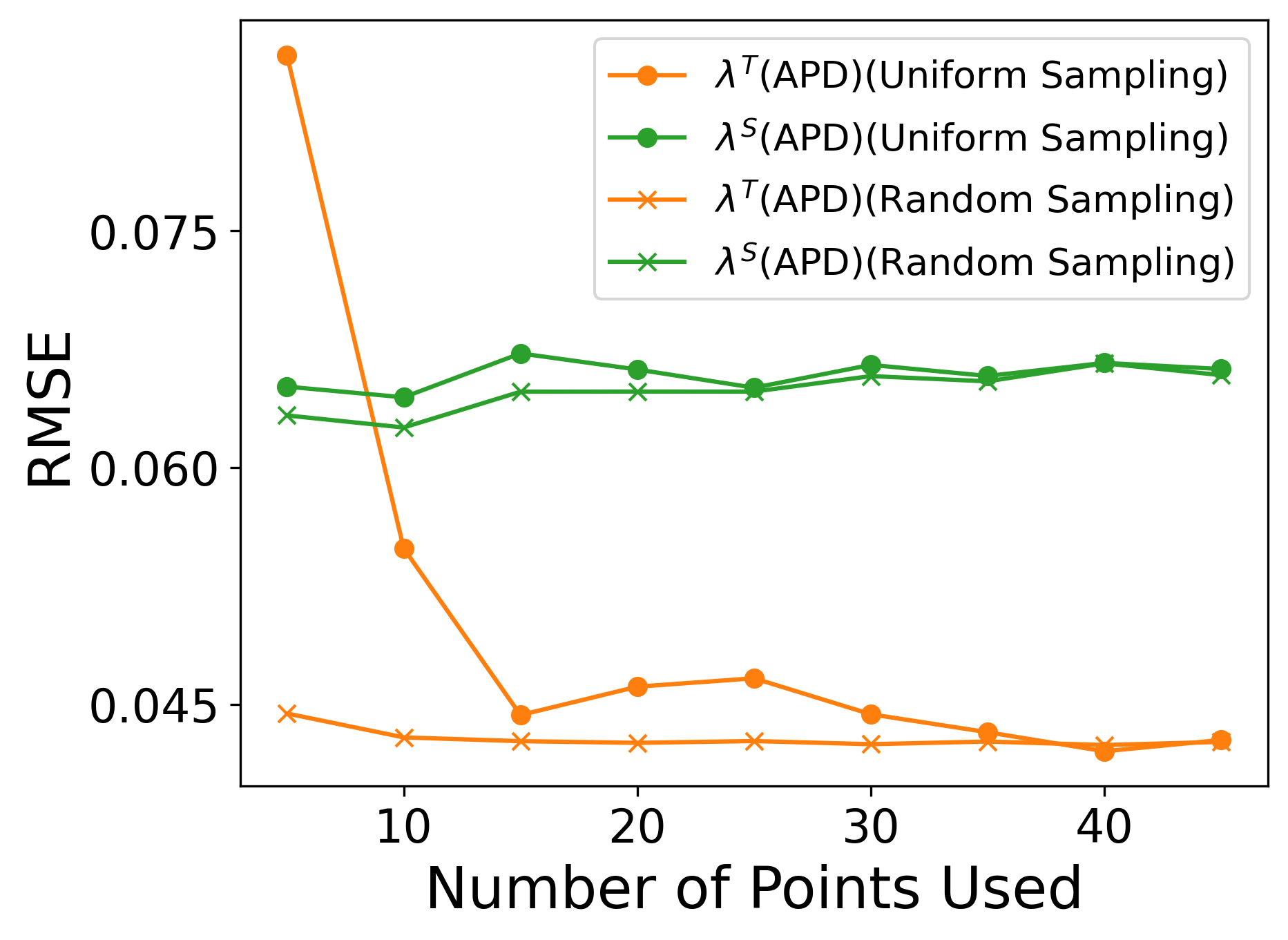}} 
    \caption{
        Root mean square error (RMSE) of LE estimations as a function of the number of spatial sampling points. Here, RMSE quantifies the error of each method compared to the one using full state variables ($\lambda^T(\text{uvw})$). %$\lambda^T(\text{APD})$ (Uniform sampling) shows a monotonic decline in RMSE as more sampling points are used, reflecting improved averaging and convergence toward the reference result. On the other hand, $\lambda^S(\text{APD})$ (Uniform sampling) remains nearly flat across sampling densities, indicating that its one-step divergence measurement, as defined in Eq.~\ref{eq 11}, stabilizes rapidly. Although the temporal APD method eventually provides higher accuracy, it requires substantially higher spatial resolution, highlighting the trade-off between the two methods on spatial resolution side. Notably, employing random sampling drastically improves the accuracy of $\lambda^T(\text{APD})$ at sparse resolutions (fewer than 15 points). Random sampling has a negligible effect on $\lambda^S(\text{APD})$.
    }
    \label{fig:exp1}
\end{figure}

Figure~\ref{fig:exp1} shows the relationship between root mean square error (RMSE, computed relative to $\lambda^T(uvw)$, taken as the reference truth) and the number of spatial points used for both APD-based methods. 
Under uniform sampling, RMSE decreases for $\lambda^T(\text{APD})$, as more spatial points are included, indicating improved robustness and gradual convergence, whereas $\lambda^S(\text{APD})$ shows little sensitivity to the number of points used. However, after convergence, $\lambda^T(\text{APD})$ achieves a lower RMSE than $\lambda^S(\text{APD})$.
This difference likely arises from the underlying algorithms. The spatial method, defined in Eq.~\ref{eq 11}, captures temporal divergence over a single step while placing greater emphasis on spatial chaos by memorizing all the spatial $<k,h>$ pairs, as shown in Eq.~\ref{eq 9}. Based on this formulation, it may be expected that a large number of spatial points would be needed to capture spatial complexity and achieve convergence. However, the results suggest that only a few points are sufficient for stable estimates, perhaps because of the pronounced spatiotemporal chaos displayed by the considered parameter regimes of the FK model in a 2D domain. In such scenarios, both spatial and temporal variations contribute to the overall chaotic dynamics. Hence, the averaging in Eq.~\ref{eq 11} jointly evaluates these components. As a result, longer APD sequences could offset the lack of spatial sampling, maintaining stability in the spatial LE estimates despite fewer points.

When spatial points are selected randomly, $\lambda^T(\text{APD})$ improves its accuracy at limited resolution, indicating that stochastic selection of measurement locations effectively mitigates the biases of sparse uniform grids and provides a more robust representation of the underlying dynamics by avoiding non-representative spatial configurations. In contrast, the $\lambda^S(\text{APD})$ algorithm's one-step divergence formulation is inherently stable and reaches rapid convergence even with minimal spatial data, making it less sensitive to the specific arrangement of sampling points.

\begin{figure}[htbp!]
    \centering
    \hspace*{-1cm}
    {\includegraphics[width=0.46\textwidth]{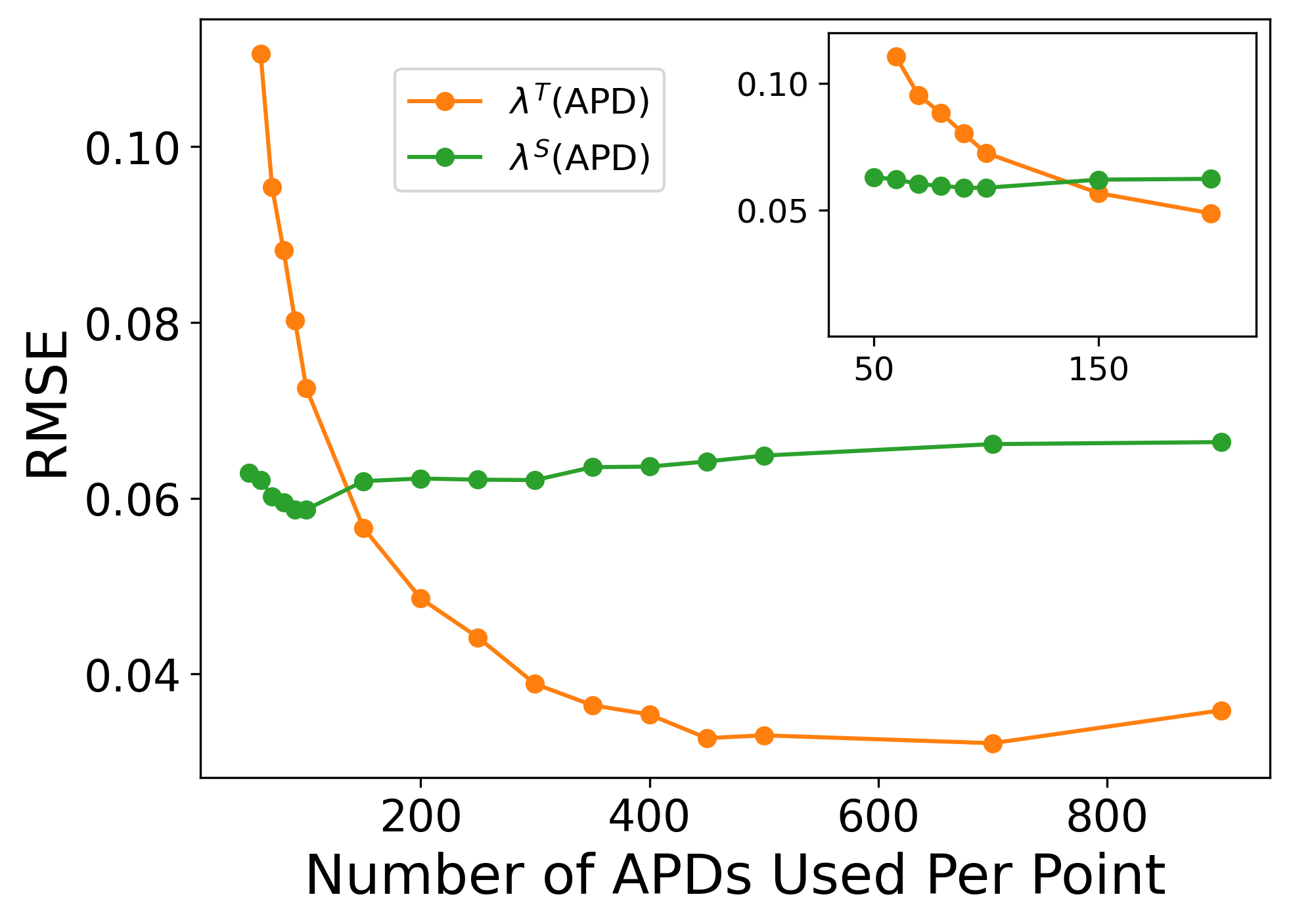}} 
\caption{
    RMSE of LE estimations plotted against number of APDs used per point. Each point, $N_{\text{APD}}$ typically ranges from [900,1300]. To evaluate the impact of sequence length on estimation accuracy, these sequences were truncated to a uniform length across all points. For example, retain only the first 300 APDs at each point before calculating the LEs. The inset provides a magnified view of the behavior for short sequences ($N_{\text{APD}} < 200$).
    %$\lambda^T(\text{APD})$ shows steady improvement with longer recordings, as extended sequences enhance averaging and convergence toward the benchmark. $\lambda^S(\text{APD})$ remains nearly constant across sequence lengths, consistent with its one-step divergence measurement in the algorithm, which requires minimal temporal data. Although $\lambda^T(\text{APD})$ ultimately provides higher accuracy, it needs longer recordings of APDs, highlighting the trade-off between the two methods on the temporal side. 
}
    \label{fig:exp2}
\end{figure}

Figure~\ref{fig:exp2} illustrates the effect of varying the number of APDs used per point. As expected, the temporal method benefits greatly from more APDs used, consistent with its role in capturing long-term divergence behavior. The spatial method, however, again shows minimal dependence on sequence length due to its one-step formulation. While this robustness allows it to obtain higher accuracy  with shorter time series, the resulting precision saturates at an RMSE around 0.062.

\begin{figure}[htbp!]
    \centering
    \hspace*{-1cm}
    {\includegraphics[width=0.46\textwidth]{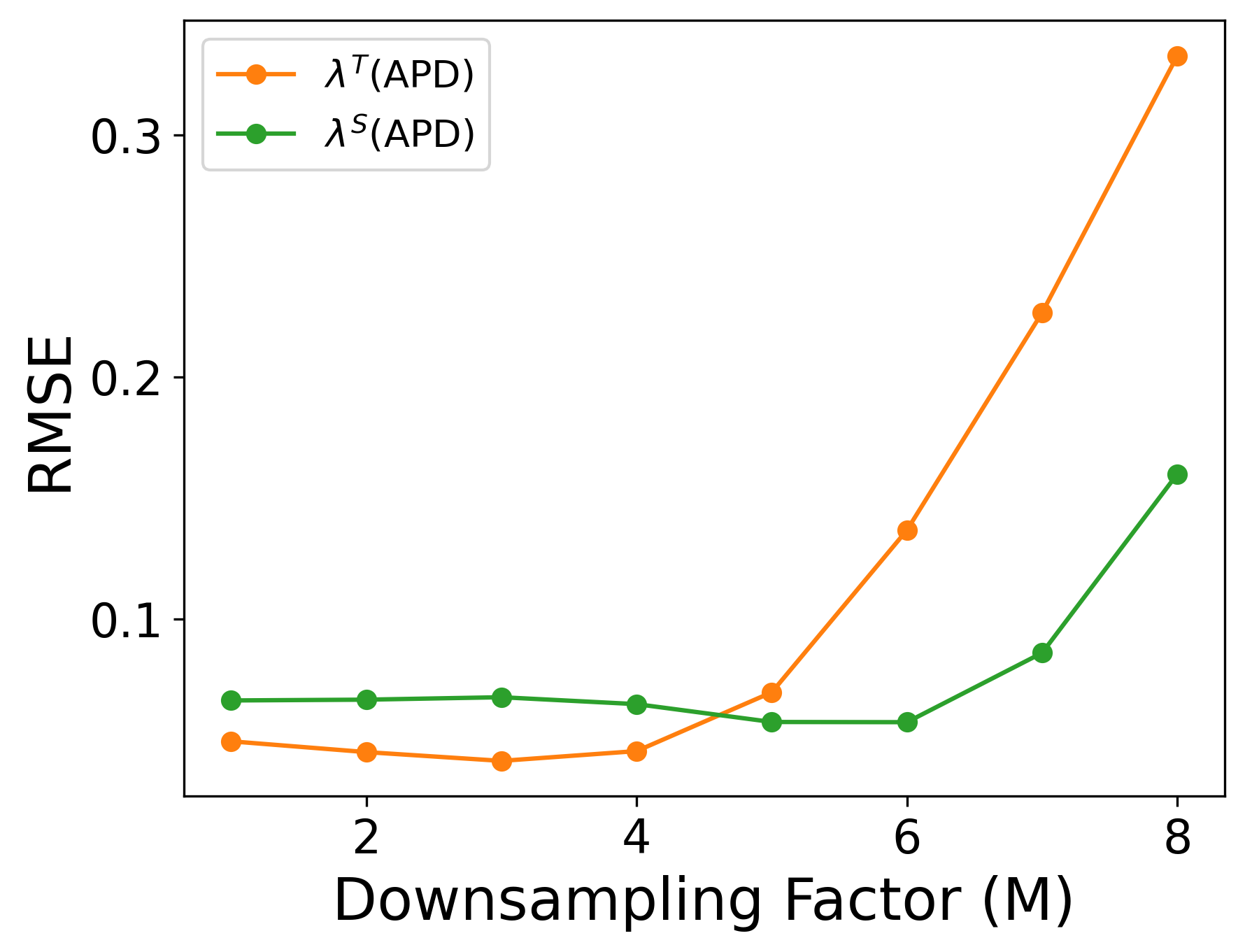}} 
\caption{
    RMSE of LE estimations plotted against downsampling factor ($M$). The adjusted measurement time step is defined as $dt' = M \cdot dt$, where $dt$ is the baseline step from Table~\ref{tab:simulation_parameters}. 
    %The error for $\lambda^T(\text{APD})$ escalates rapidly after $M = 5$, whereas $\lambda^S(\text{APD})$ maintains its stability until $M = 6$.
}
    \label{fig:exp3}
\end{figure}

We further investigated the influence of temporal resolution on estimation accuracy by applying a downsampling factor $M$ to the measurement time step $dt$. As shown in Figure~\ref{fig:exp3}, in general, estimation quality decreases with increasing $M$, with the critical threshold, beyond which the error grows, occurring at $M = 4$ for $\lambda^T(\text{APD})$ and  $M = 6$ for $\lambda^S(\text{APD})$.

\begin{figure}[htbp!]
    \centering
    \hspace*{-1cm}    
    {\includegraphics[width=0.46\textwidth]{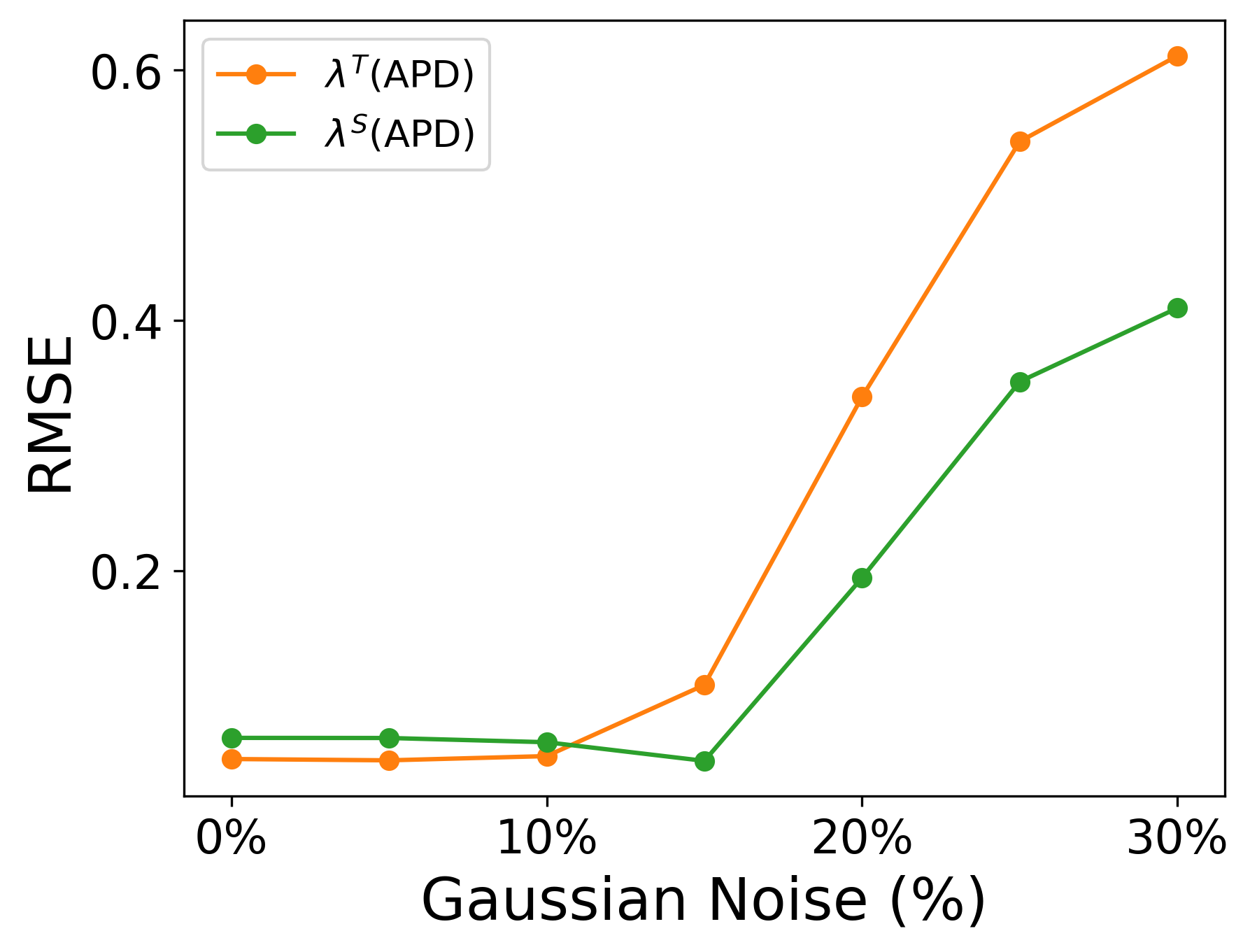}} 
\caption{
    RMSE of LE estimations plotted against added Guassian noise percentage. 
    %While both methods perform comparably at lower noise levels (up to 10\%), $\lambda^S(\text{APD})$ demonstrates higher reliability and robustness against larger noise levels compared to $\lambda^T(\text{APD})$.
}
    \label{fig:exp4}
\end{figure}

To further evaluate the robustness of the APD-based LE estimation methods under imperfect measurement conditions, we analyzed the effect of measurement noise. Figure~\ref{fig:exp4} illustrates RMSE as a function of added Gaussian noise. At low noise levels (between 0\% and 10\%), both methods demonstrate high accuracy with similarly low error rates. However, as noise exceeds 15\%, the error for the temporal method grows more rapidly than the error for teh spatial method, ultimately reaching an RMSE above 0.6 at 30\% noise compared to 0.4 for the spatial method. Because $\lambda^S(\text{APD})$ leverages spatial correlations across the tissue, it inherently filters out localized, independent random fluctuations. Consequently, it maintains a lower RMSE, confirming that the spatial APD method is advantageous for applications where elevated noise levels are unavoidable.

%These experiments reveal a distinct trade-off between the two methods. The temporal approach is data-intensive but capable of high precision when an adequate and noise-free input is available. The spatial method proves to be highly data-efficient and robust. It successfully characterizes complex spatiotemporal chaos, remains stable during the short measurement windows of high-dimensional models (such as the TNNP model), and exhibits high robustness against Gaussian noise. This distinction is particularly relevant for experimental applications where spatial resolution or temporal duration is limited.

\section{Discussion and Conclusion}

We introduced and evaluated two APD-based methods for estimating LE to characterize spatiotemporal chaos in cardiac tissue models. Both approaches operate on reconstructed data from APD and rely on lag embedding to recover underlying dynamics. The temporal method emphasizes long-range divergence, using spatial information for averaging. The spatial method, on the other hand, evaluates one-step divergence between neighboring spatial points at each time frame, with temporal length contributing to statistical averaging. Together, they provide complementary ways to assess dynamical instability from reduced data.

To assess performance, we compared these methods against the standard method by utlizing full state variables locally with Wolf's algorithm. We have conducted comparsions on single spirals and multilple spirals cases and both algorithms perform decent precision in LE estimations.

We examined how the two methods reflect the structure of spiral wave patterns. In relatively stationary single-spiral cases, such as cases A, E, and F in Fig.~\ref{fig: single_spiral}, both $\lambda^T$ and $\lambda^S$ align well with reference values, confirming their ability to capture periodic dynamics or low-dimensional chaos. In cases with meandering or drifting cores (e.g., cases B–D in Fig.~\ref{fig: single_spiral} and D2 in Fig.~\ref{fig:D2}), the methods diverge in their results: $\lambda^T$ detects chaos when spiral cores undergo displacement, indicating temporal instability at the pixel level, while $\lambda^S$ is more sensitive to the geometric irregularity of the spiral shape itself. This is particularly evident in case D2, where the spiral cores remain stationary but exhibit distorted, non-periodic shapes, resulting in a periodic signal under the temporal method but a chaotic signature under the spatial method. A similar pattern appears in Fig.~\ref{fig:heat map}, where red regions in $\lambda^T$ map correspond to drifting cores (e.g., B and E), while spatial irregularity without displacement leads to predominantly blue (periodic) signals in the temporal map, despite the spatial LE indicating local instability.

In more disordered multi-spiral regimes, we investigated how tissue instability responds to ionic perturbations by systematically varying parameters associated with gating variables in both the FK and TNNP models. Across different tuning directions, both APD-based LE estimators revealed clear sensitivity to gating kinetics, with changes in certain parameters either increasing or decreasing the chaosity. These outcomes reflect how specific ionic time constants modulate recovery dynamics and, in turn, influence the prevalence of wavebreaks and reentrant activity. This supports the broader view that spatiotemporal complexity in excitable media is tightly linked to the interplay between recovery properties and wavefront propagation. Furthermore, We observe that the underlying dynamics are directly encoded in the system's spatial structure, where the progression from coherent to fragmented spiral core trajectories mirrors the shift in LE values. Such geometric transitions suggest the potential for developing image-based chaos classification methods. In particular, convolutional neural network (CNN)\cite{lecun2002gradient}, commonly used for pattern recognition, may be employed to learn and classify levels of chaoticity based on spatial features of spiral core evolution.

Finally, we analyzed the sensitivity of APD-based methods to data quality by varying spatial sampling density, APD sequence length, downsampling resolution, and noise levels to define baseline criteria. $\lambda^{T}(\text{APD})$ proved highly sensitive to these constraints, requiring at least 15 spatial points in a tissue with length 18~cm, as detailed in Tab.~\ref{tab:simulation_parameters} to retain robustness, although this requirement can be mitigated if random sampling is used instead of a uniform grid. Furthermore, $\lambda^{T}(\text{APD})$ demands longer APD sequences to reliably capture long-term divergence and remains accurate only within a narrow temporal resolution (downsampling factor $M \leq 4$) and low-noise environments where Gaussian noise is below 15\%. Conversely, by incorporating the spatial information to algorithm, $\lambda^{S}(\text{APD})$ provides a significantly more robust alternative for constrained datasets. It stabilizes rapidly even with minimal sampling points and short sequence lengths due to its one-step divergence formulation. This resilience extends to a higher downsampling threshold of $M=6$ and noise levels as high as 20\%, suggesting that $\lambda^{S}(\text{APD})$ has the potential to be the primary tool for clinical or experimental mapping where recordings are inherently sparse, short, and noisy.

In summary, the temporal method offers higher accuracy and better sensitivity to dynamic transitions but requires more data. The spatial method is computationally efficient, robust to limited sampling, and performs reliably in regimes with strong spatiotemporal coupling. Depending on the constraints, one can prioritize the method that balances accuracy with available data. Together, these methods provide a practical toolkit for assessing chaos in cardiac systems using accessible observables.

\section{Limitations and Future Work}

While this study demonstrates the efficacy of applying APD-based LE estimation in cardiac models, several limitations persist that suggest key directions for future research. The current analysis relies on simulation data from ionic models designed to replicate human and animal electrophysiology. Our noise testing was restricted to synthetic Gaussian noise added to ideal voltage data. In real-world clinical and experimental environments, data obtained via optical mapping or ECGs often contain biological and measurement noise, which cannot be emulated purely by adding Gaussian noise to voltage \cite{laughner2012processing,xie2020computational,zgallai2013characterization}. This noise could potentially compromise the accuracy of phase-space reconstruction and the final LE estimation. Future research should focus on applying these algorithms to experimental and clinical datasets to ensure their functionality and robustness when dealing with real-world recordings.

Furthermore, the simulations conducted in this study were restricted to 2D tissue, whilst the real heart is a 3D structure, where transmural filaments and scroll waves contribute significantly to the complexity of spatiotemporal chaos. Future validation should extend the APD-based LE framework to 3D cardiac models to determine if limited observations can effectively capture instabilities in deep tissue.

Beyond extending the study to more complex systems, the ability to quantify dynamical complexity allows our method to be integrated into machine learning pipelines as a core feature for prediction. This integration could facilitate advanced forecasting of future system states, including time-series evolution, complexity transitions, and fatality risks. By successfully quantifying chaos under different parameter regimes, our research provides a Possible roadmap for chaos control, offering insights into steering a system from lethal chaotic dynamics toward stable periodic rhythms. Moreover, the spatial patterns revealed by LE distributions may serve as indicators of vulnerable regions in the tissue, supporting classification and control strategies in arrhythmia management.

\section*{Code and data availability}
The complete simulation and analysis code, including all WebGL applications and pre-calculated data, is publicly available at \href{https://github.com/William-XiaodongAn/Quantifying-The-Complex-Spatiotemporal-Chaos-of-Cardiac-Fibrillation-in-Ionic-Models}{https://github.com/William-XiaodongAn/Quantifying-The-Complex-Spatiotemporal-Chaos-of-Cardiac-Fibrillation-in-Ionic-Models}.

\section*{Acknowledgments}
We thank [names of colleagues] for valuable discussions and [any lab members] for technical assistance. 
This work was supported by [funding agency and grant number if applicable]. 
Computations were performed using resources provided by [institution0].

\appendix

\section{Results For $\tau_0$ and $\tau_{si}$}

\begin{figure}
    \centering
    {\includegraphics[width=0.48\textwidth]{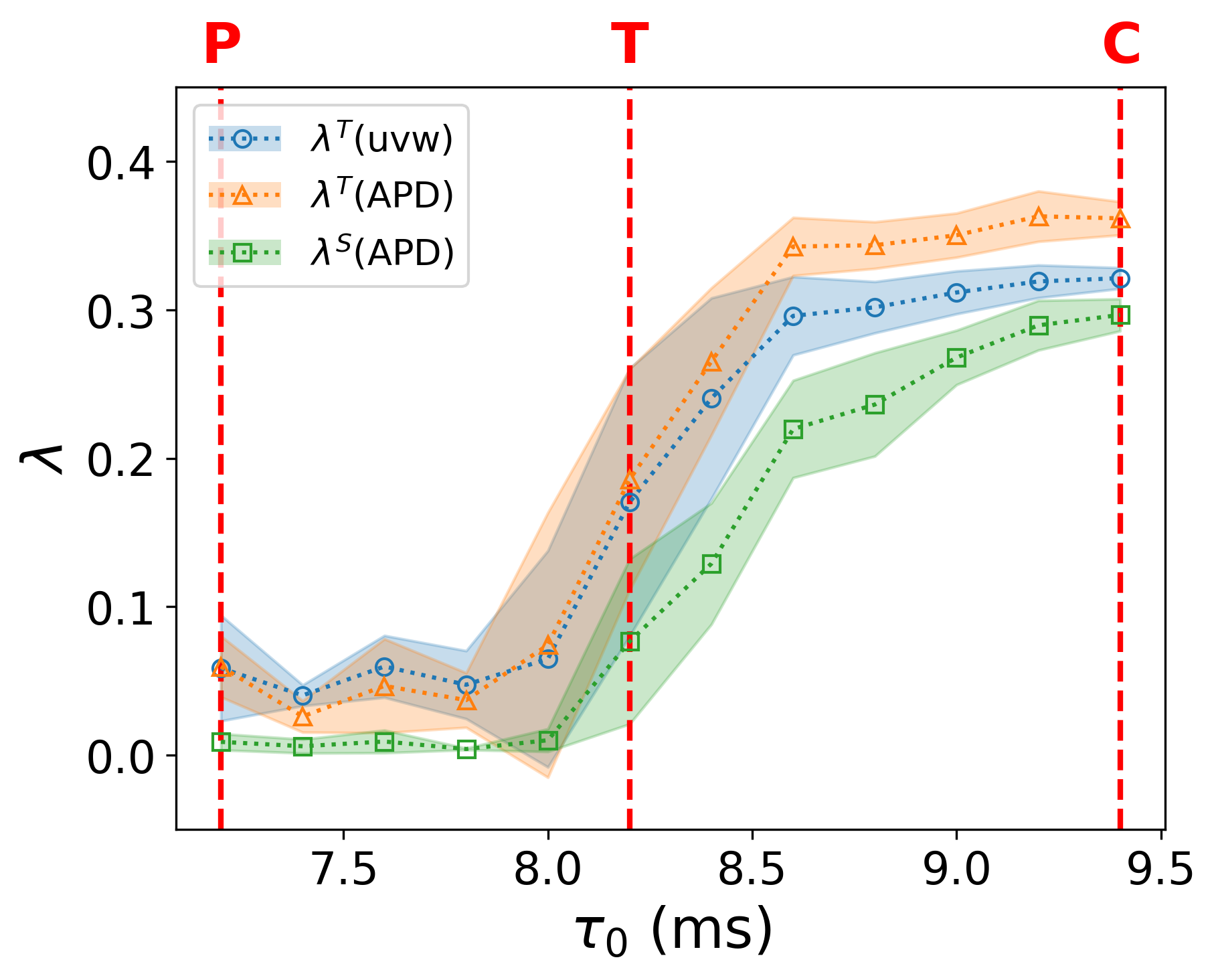}} \\
    {\includegraphics[width=0.5\textwidth]{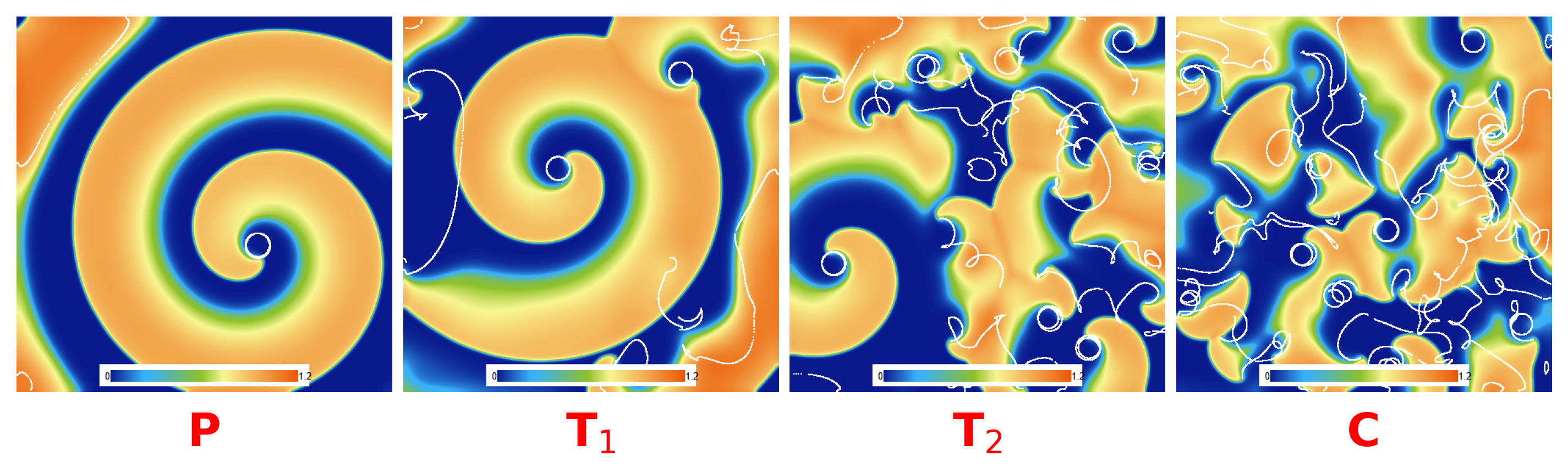}} \\
    {\includegraphics[width=0.45\textwidth]{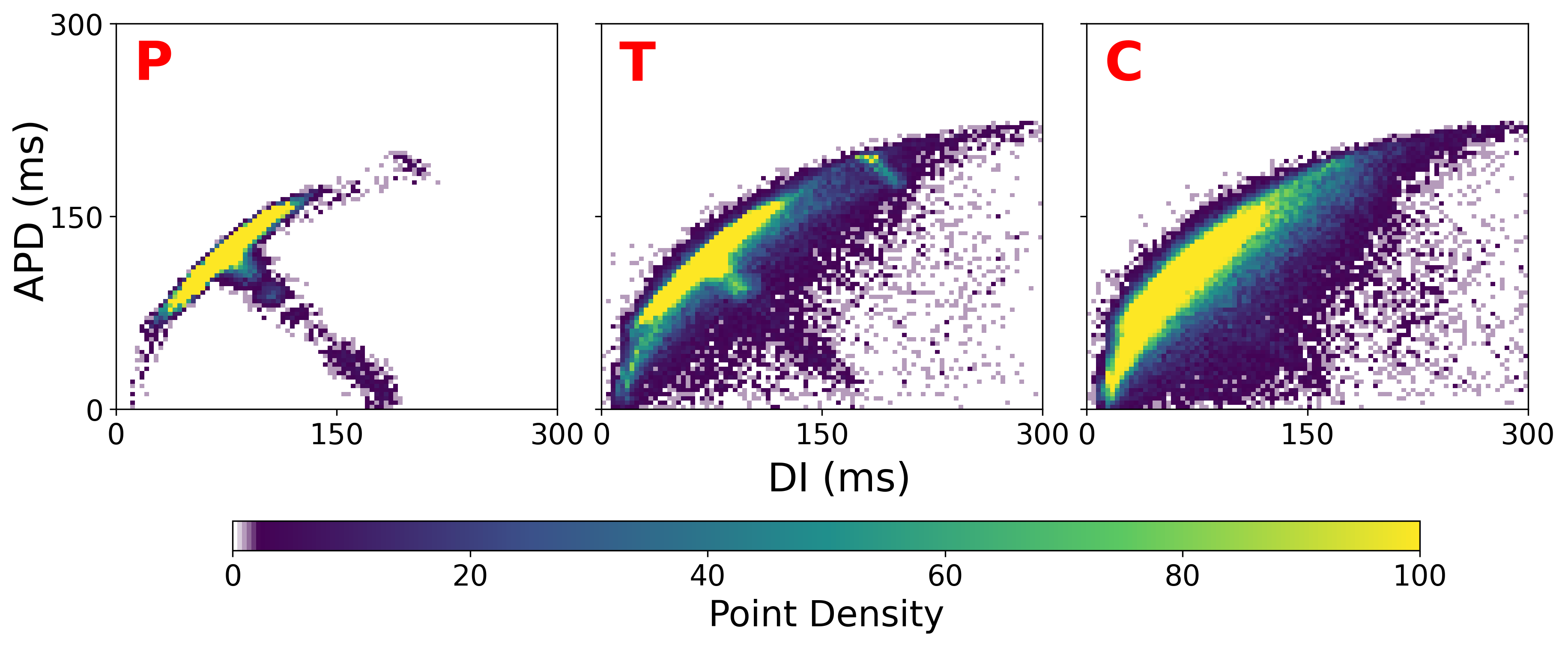}} 
\caption{Performance of LE estimates for different spiral wave dynamics obtained by varying $\tau_0$ in the FK model. 
\textbf{Top:} LEs from state variables and APD signals, as in Fig.~\ref{fig:taud}.
Labeled points correspond to example values of $\tau_0$ within the periodic (A), transitional (B), and chaotic (C) regimes. 
\textbf{Middle:} Simulation snapshots with spiral tip trajectories (white). 
\textbf{Bottom:} Restitution curves of APD versus DI for the different states, plotted with range [0,300] on both axes.}

    \label{fig:tau0}
\end{figure}

\begin{figure}
    \centering
    {\includegraphics[width=0.48\textwidth]{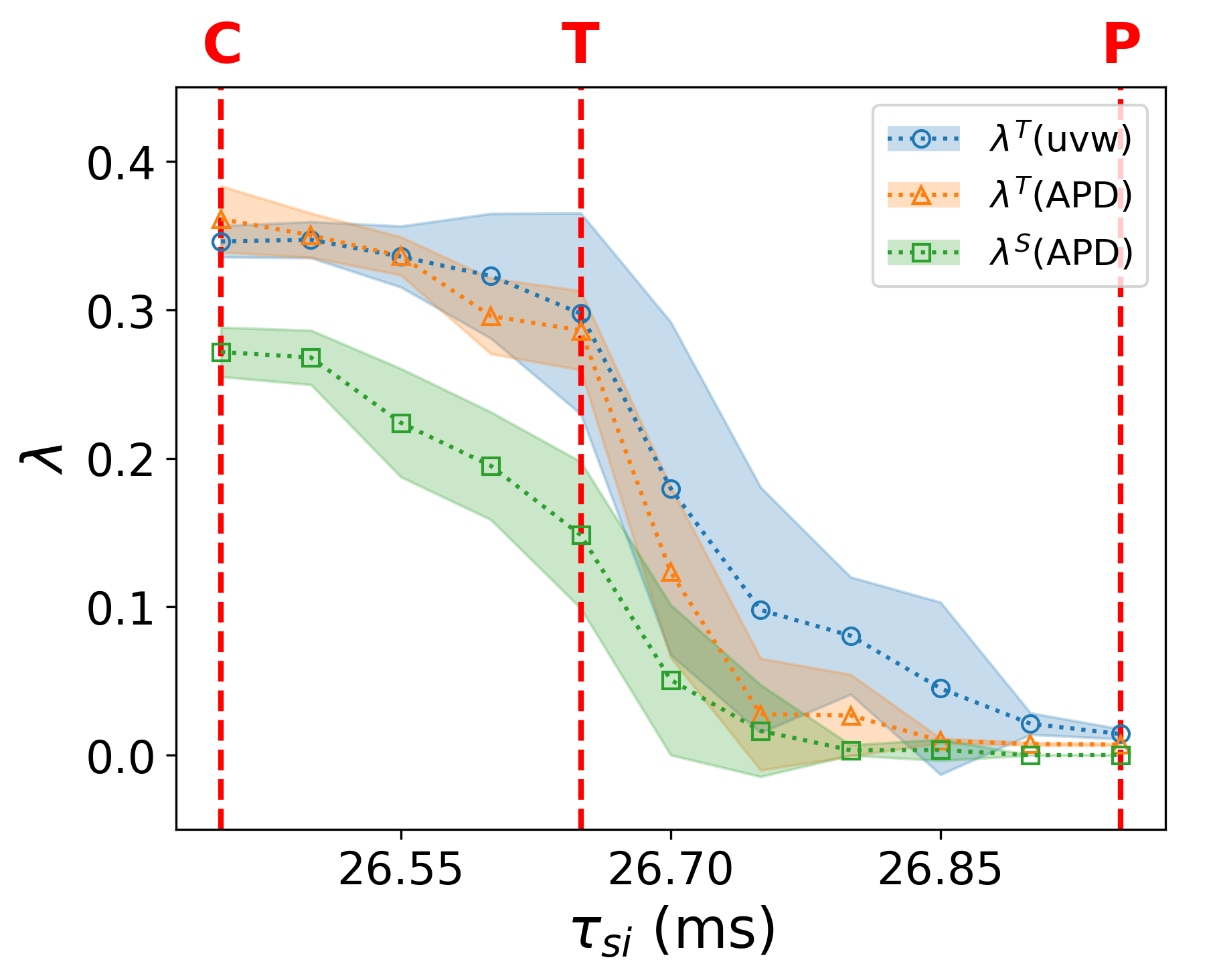}} \\
    {\includegraphics[width=0.5\textwidth]{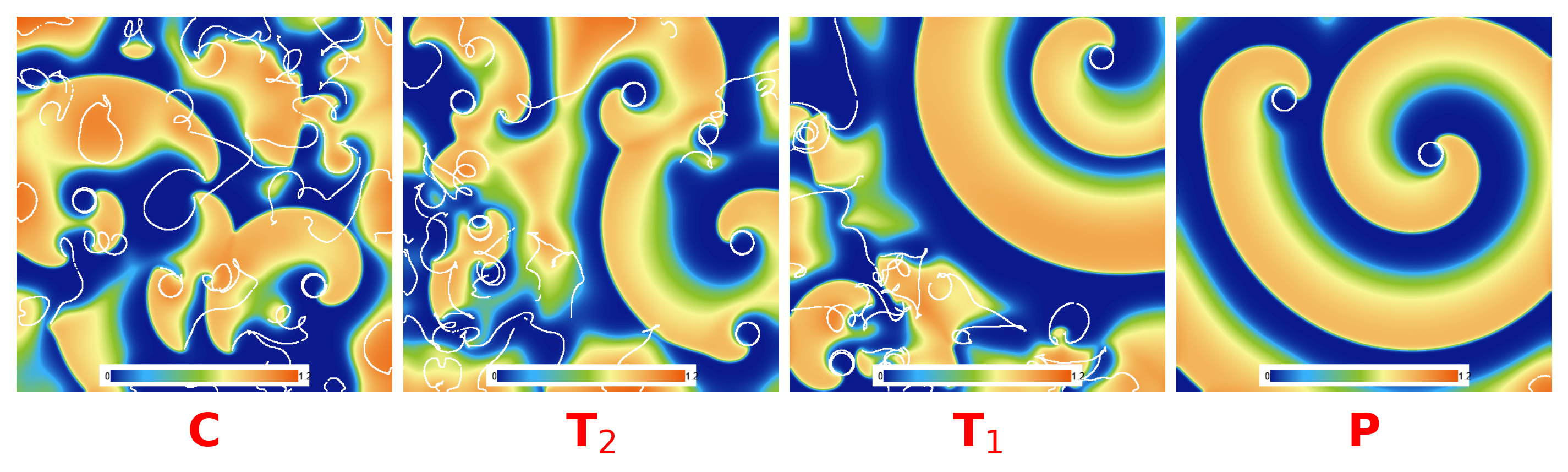}} \\
    {\includegraphics[width=0.45\textwidth]{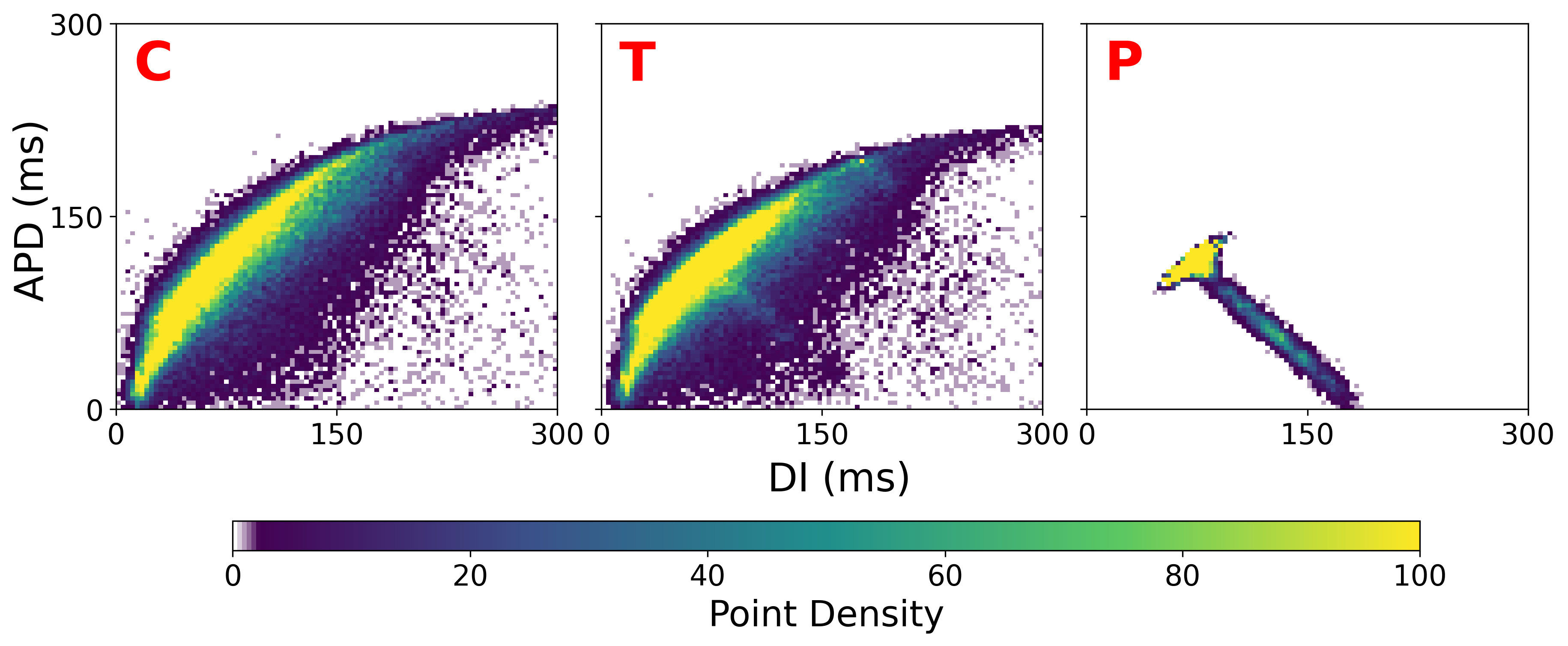}} 
\caption{Performance of LE estimates for different spiral wave dynamics obtained by varying $\tau_{si}$ in the FK model. 
\textbf{Top:} LEs from state variables and APD signals, as in Fig.~\ref{fig:taud}.
Labeled points correspond to example values of $\tau_{si}$ within the chaotic (A), transitional (B), and periodic (C) regimes. 
\textbf{Middle:} Simulation snapshots with spiral tip trajectories (white). The transitional state exhibits mixed quasiperiodic and chaotic features. 
\textbf{Bottom:} Restitution curves of APD versus DI for the three regimes, plotted with range [0,300] on both axes.}
\label{fig:tausi}
\end{figure}
The $\tau_0$ dataset shown in Fig.~\ref{fig:tau0} demonstrates a significantly reduced change rate of the LE. The transition window spans a wider range at approximately [7, 9.5], which differs by several orders of magnitude compared to the sharp transitions observed in $\tau_d$ and $\tau_r$. This suggests that $\tau_0$ plays a more stabilizing role, yielding a slower and smoother progression from chaos to period.

Fig.~\ref{fig:tausi} illustrates the behavior under variations of $\tau_{si}$. Once again, a similar transition state $B$ emerges where quasiperiodic and periodic behaviors overlap. Additionally, the directionality of LE change appears parameter-dependent. For parameters such as $\tau_0$ and $\tau_r$, an increase tends to widen the spiral wave, which then requires a larger domain to maintain wave coherence. That is to say, when constrained to a fixed spatial domain, this leads to more frequent wave collisions and breakups, indicating a higher LE and stronger chaos. This trend reinforces the notion that parameter-induced geometric alterations in spiral waves are central to chaos modulation.

\bibliographystyle{apsrev4-2}
\bibliography{aipsamp}
\end{document}